\documentclass[smallextended]{svjour3}
\usepackage[style=ieee,backend=biber,sorting=nyt]{biblatex}
\usepackage{caption}
\usepackage{subcaption}
\usepackage{graphicx}
\usepackage{amsmath}
\usepackage{amssymb}
\usepackage{multirow}
\usepackage{caption}

\usepackage[dvipsnames,table,xcdraw]{xcolor}
\usepackage{soul}
\usepackage{url}
\usepackage[hidelinks]{hyperref}
\usepackage{booktabs}
\usepackage{multirow}
\usepackage{xspace}
\usepackage{float}
\usepackage{tcolorbox}
\usepackage{xcolor}
\usepackage{listings}
\usepackage{appendix}
\usepackage{balance}
\newcommand{\canvas}{\texttt{<canvas>}\xspace}
\newcommand{\rqone}{To what extent can VLMs detect visual bugs in HTML5 \canvas applications?}
\newcommand{\rqtwo}{How does providing additional text and image context impact the capabilities of VLMs to detect visual bugs in HTML5 \canvas applications?}
\newcommand{\State}{\textit{State}\xspace}
\newcommand{\Rendering}{\textit{Rendering}\xspace}
\newcommand{\Layout}{\textit{Layout}\xspace}
\newcommand{\Appearance}{\textit{Appearance}\xspace}
\newcommand{\proceduralgraphics}{\textit{Procedural}\xspace}
\newcommand{\assetbasedgraphics}{\textit{Asset-Based}\xspace}
\newcommand{\animations}{\textit{Animations}\xspace}
\newcommand{\game}{\textit{Game}\xspace}
\newcommand{\datavisualization}{\textit{Data Visualization}\xspace}
\newcommand{\visualeditor}{\textit{Visual Editor}\xspace}
\newcommand{\gptfouro}{\mbox{GPT-4o}\xspace}
\newcommand{\bugfree}{\mbox{bug-free}\xspace}
\newcommand{\buginjected}{\mbox{bug-injected}\xspace}
\newcommand{\promptzero}{\textit{NoContext}\xspace}
\newcommand{\promptone}{\textit{README}\xspace}
\newcommand{\prompttwo}{\textit{README+BugDescriptions}\xspace}
\newcommand{\promptthree}{\textit{AllContextExceptAssets}\xspace}
\newcommand{\promptfour}{\textit{AllContext}\xspace}
\newcommand{\promptablationA}{\textit{README(Good)}\xspace}
\newcommand{\promptablationB}{\textit{README(Bad)}\xspace}
\newcommand{\Precision}{precision\xspace}
\newcommand{\Recall}{recall\xspace}
\newcommand{\PixiJS}{\textsc{PixiJS}\xspace}
\newcommand{\ThreeJs}{\textsc{ThreeJS}\xspace}
\definecolor{lightgrey}{rgb}{0.9, 0.9, 0.9} 
\definecolor{backcolour}{rgb}{0.97,0.97,0.97} 
\definecolor{darkgray}{rgb}{0.4,0.4,0.4}      
\definecolor{bluegray}{rgb}{0.2,0.4,0.6}      
\definecolor{darkgreen}{rgb}{0,0.5,0}         
\definecolor{darkpurple}{rgb}{0.4,0.2,0.6}    
\definecolor{black}{rgb}{0,0,0}               
\lstdefinestyle{Markdown}{
    backgroundcolor=\color{backcolour},       
    commentstyle=\color{darkgreen}\itshape,  
    keywordstyle=\color{bluegray}\bfseries,  
    numberstyle=\tiny\color{darkgray},       
    stringstyle=\color{darkpurple},          
    basicstyle=\ttfamily\small\color{black}, 
    breakatwhitespace=false,                 
    breaklines=true,                         
    keepspaces=true,                         
    numbers=left,                            
    numbersep=5pt,                           
    showspaces=false,                        
    showstringspaces=false,                  
    showtabs=false,                          
    tabsize=2,                               
    literate={\`}{\textasciigrave}1          
}
\makeatletter
\let\blx@rerun@biber\relax
\makeatother

\begin{document}
\title{Exploring the Capabilities of Vision-Language Models to Detect Visual Bugs in HTML5 \canvas Applications}
\titlerunning{Detecting Visual Bugs in HTML5 \canvas applications}

\author{Finlay~Macklon, Cor-Paul~Bezemer} 

\institute{Finlay~Macklon \and Cor-Paul~Bezemer \at
              Analytics of Software, Games and Repository Data (ASGAARD) Lab \\
              University of Alberta \\
              Edmonton, AB, Canada \\
              \email{\{macklon,bezemer\}@ualberta.ca}}

\date{Received: date / Accepted: date}

\maketitle

\begin{abstract}
The HyperText Markup Language 5 (HTML5) \canvas is useful for creating visual-centric web applications.
However, unlike traditional web applications, HTML5 \canvas applications render objects onto the \canvas bitmap without representing them in the Document Object Model (DOM).
Mismatches between the expected and actual visual output of the \canvas bitmap are termed visual bugs. 
Due to the visual-centric nature of \canvas applications, visual bugs are important to detect because such bugs can render a \canvas application useless.
As we showed in prior work, \assetbasedgraphics graphics can provide the ground truth for a visual test oracle. 
However, many \canvas applications procedurally generate their graphics.
In this paper, we investigate how to detect visual bugs in \canvas applications that use \proceduralgraphics graphics as well.
In particular, we explore the potential of Vision-Language Models (VLMs) to automatically detect visual bugs.
Instead of defining an exact visual test oracle, information about the application’s expected functionality (the context) can be provided with the screenshot as input to the VLM.
To evaluate this approach, we constructed a dataset containing 80 bug-injected screenshots across four visual bug types (\Layout, \Rendering, \Appearance, and \State) plus 20 bug-free screenshots from 20 \canvas applications.
We ran experiments with a state-of-the-art VLM using several combinations of text and image context to describe each application’s expected functionality.
Our results show that by providing the application README(s), a description of visual bug types, and a bug-free screenshot as context, VLMs can be leveraged to detect visual bugs with up to 100\% per-application accuracy.
\end{abstract}


\section{Introduction}\label{sec:introduction}
The HyperText Markup Language 5 (HTML5) \canvas is used to develop visual-centric web applications such as games and data visualizations~\cite{macklon2023taxonomy, applecanvas}.
HTML5 \canvas applications can be created entirely within the HTML \canvas element by attaching listeners to the browser window for events such as mouse clicks~\cite{macklon2023taxonomy}.
When a browser event listener is triggered, custom scripts that leverage the Canvas API or WebGL API can be used to update the contents of the \canvas.
However, unlike traditional web applications, the contents of the \canvas are rendered as a bitmap and are not represented in the Document Object Model (DOM).
Existing web testing tools that typically rely on analysis of the DOM to drive test automation are very limited for testing the \canvas~\cite{macklon2023taxonomy, bajammal2018web}, necessitating new testing approaches and tools that test the \canvas.
Mismatches between the expected and actual visual output of the \canvas bitmap are termed visual bugs, and we found in our prior work~\cite{macklon2023taxonomy} that visual bugs are of particular concern when developing \canvas applications (e.g., because they may render a \canvas application unusable).

Visual objects can be rendered on the \canvas using: 
\begin{enumerate}
    \item  \assetbasedgraphics graphics, which depend on image assets, or 
    \item \proceduralgraphics graphics that are generated on-the-fly.
\end{enumerate}
In our prior work, we proposed a testing approach to target visual bugs in \canvas applications that use \assetbasedgraphics graphics~\cite{macklon2022automatically}.
However, the approach described in our prior work requires that the \canvas application has a \canvas objects representation (COR), and specifying visual test oracles is challenging due to the creation and maintenance of the complex image processing algorithms.
Meanwhile, no testing approaches exist for \canvas applications that use \proceduralgraphics graphics.
As a result, the testing of \canvas applications is mostly done manually~\cite{viggiato2022identifying, viggiato2022using, viggiato2023prioritizing}.

Rather than maintaining visual test oracles that require the specification of exactly what we are (not) looking for, in this paper we aim to avoid the oracle problem by leveraging Vision-Language Models (VLMs).
VLMs are large, pre-trained deep learning models that process image and text inputs simultaneously~\cite{zhang2024vision}.
We hypothesize that VLMs could be leveraged to automatically analyze a screenshot of a \canvas application to determine if a visual bug is displayed on the \canvas bitmap, without specifying a visual test oracle.
A similar approach has been proven successful for open-world video games~\cite{taesiri2024glitchbench}, which are visual-centric like \canvas applications but often photorealistic.
Photorealistic images are well represented in a VLM's pre-training dataset that contains large amounts of real-world images available on the internet~\cite{zhang2024vision}.
However, HTML5 \canvas applications are often not photorealistic.
Therefore, we also hypothesize that it is more difficult to detect visual bugs using a VLM in \canvas applications than open-world games because the VLM has not been trained on the necessary information to describe what the \canvas bitmap should look like.

In this paper, we address the following research questions:

\begin{itemize}
    \item RQ1: \emph{\rqone}
    \item RQ2: \emph{\rqtwo}
\end{itemize}

\begin{figure*}[htbp]
\centering

\setlength{\fboxsep}{0pt}%
\begin{subfigure}[t]{\linewidth}
\centering
\fbox{\includegraphics[width=\textwidth]{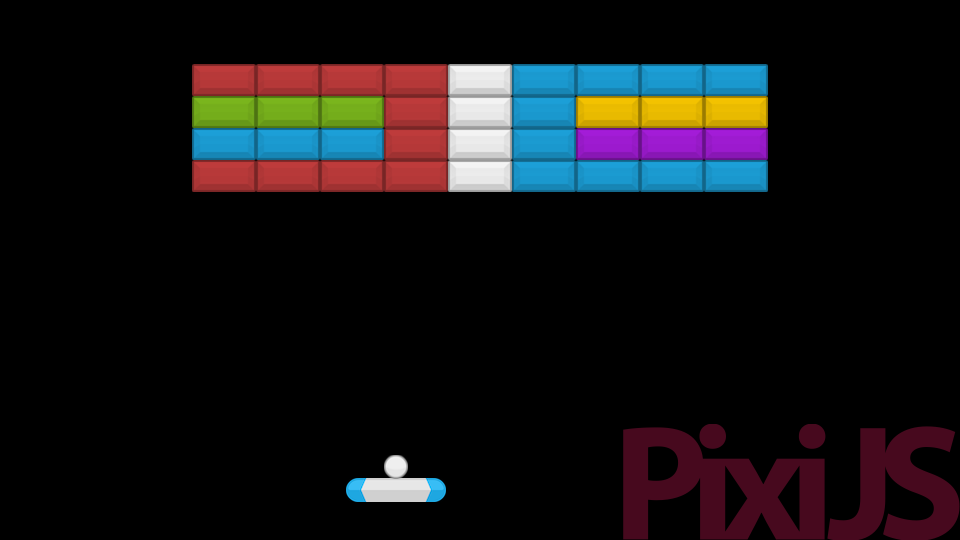}}
\end{subfigure}

\setlength{\fboxsep}{8.5pt}%
\begin{subfigure}[t]{\linewidth}
\fcolorbox{black}{lightgrey}{%
\parbox{0.95\columnwidth}{
{\footnotesize
\textbf{Detected a Visual Bug?} No
\\
\textbf{Description of Visual Bug:} \textit{``''}
\\
\textbf{Correct Answer?} Yes \checkmark
}}}
\caption{Screenshot without any visual bugs.}
\label{fig:intro_a}
\end{subfigure}

\setlength{\fboxsep}{0pt}%
\begin{subfigure}[t]{\linewidth}
\centering
\fbox{\includegraphics[width=\textwidth]{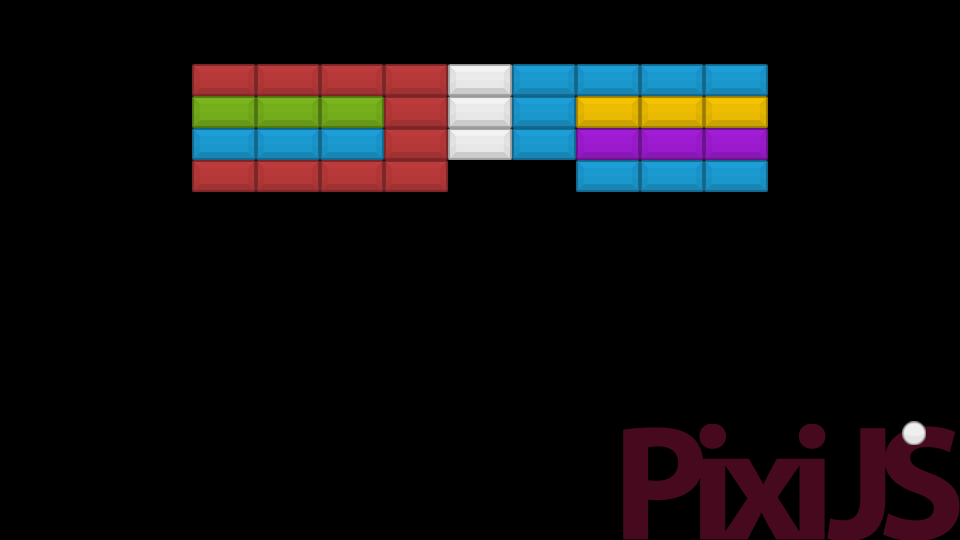}}
\end{subfigure}

\setlength{\fboxsep}{8.5pt}%
\begin{subfigure}[t]{\linewidth}
\fcolorbox{black}{lightgrey}{%
\parbox{0.95\columnwidth}{
{\footnotesize
\textbf{Detected a Visual Bug?}  Yes
\\
\textbf{Description of Visual Bug:} \textit{``The paddle, which should be present at the bottom of the screen to interact with the ball, is missing. This could be an issue related to object state or layout, where the paddle is not visible or has been positioned incorrectly.''}
\\
\textbf{Correct Answer?} Yes \checkmark
}}}
\caption{Screenshot with a \textit{State} bug making the paddle invisible.}
\label{fig:intro_b}
\end{subfigure}

\caption{Two screenshots from a Breakthrough clone (\textsc{ourcade/ecs-dependency-injection} in Table~\ref{tab:subsetcanvasapplications}) coupled with the (correct) visual bug detection results generated with \gptfouro while utilizing prompting strategy \promptthree from Table~\ref{tab:promptstrategies}.}
\label{fig:intro_sample}
\end{figure*}

We evaluate the capabilities of VLMs to detect four types of visual bugs (\Layout, \Rendering, \Appearance, and \State)~\cite{macklon2023taxonomy} across 80 bug-injected screenshots plus 20 bug-free screenshots from 20 free and open-source (FOSS) HTML5 \canvas applications.
We use several combinations of text and image context to describe the correct functionality of the application while prompting the VLM (referred to as ``prompting strategies'' in our paper). Some of the inputs we use are application README file(s), a description of potential visual bugs, and a \bugfree screenshot.
Figure~\ref{fig:intro_sample} shows one \bugfree screenshot and one \buginjected screenshot each paired with the bug detection results generated by a VLM.
We find that the best overall prompting strategy (\promptthree in Table~\ref{tab:promptstrategies}) yields up to 100\% accuracy for a \canvas application, but also find that accuracy can vary widely depending upon the application under test.

The main contributions of our paper are as follows:

\begin{itemize}
    \item Empirical results indicating that VLMs are useful tools for detecting visual bugs in HTML5 \canvas applications.
    \item A dataset of 100 screenshots\footnote{\url{https://bit.ly/vlm_canvas_bugs-data}} from 20 FOSS HTML5 \canvas applications, consisting of 20 \bugfree screenshots and 80 \buginjected screenshots across four visual bug types.    
    \item A framework\footnote{\url{https://bit.ly/vlm_canvas_bugs-code}} for collecting bug-free and bug-injected screenshots of HTML5 \canvas applications built with \PixiJS to support future research on HTML5 \canvas testing.
\end{itemize}

Our paper is structured as follows.
Section~\ref{sec:background} provides background information to motivate our study.
Section~\ref{sec:related} discusses related work.
Section~\ref{sec:leveragingvlms} describes how we leverage VLMs to detect visual bugs.
Section~\ref{sec:methodology} presents our methodology for constructing our dataset.
Section~\ref{sec:experiments} presents the design of our experiments.
Section~\ref{sec:results} presents our results and discusses our findings for each of our research questions.
Section~\ref{sec:threats} addresses threats to validity.
Section~\ref{sec:conclusion} concludes our paper.

\section{Background} \label{sec:background}
In this section we provide background information on HTML5 \canvas applications, the challenges associated with detecting visual bugs, and why VLMs offer a potential solution to some of the challenges posed when testing HTML5 \canvas applications.

\subsection{HTML5 \canvas Applications}~\label{subsec:html5canvasapplications}
The HTML5 \canvas enables developers to create interactive web applications entirely inside of the HTML \canvas element~\cite{macklon2023taxonomy, macklon2022automatically, ibmcanvas}.
The HTML \canvas element renders a bitmap that is updated through function calls by the Canvas Application Programming Interface (API)~\cite{mozillacanvasapitutorial} or the WebGL API~\cite{mozillawebglapitutorial}.
Web developers often face implementation issues when utilizing the APIs for the \canvas~\cite{bajaj2014mining}, so instead developers may use \canvas rendering frameworks~\cite{macklon2022automatically} such as \PixiJS or \ThreeJs.
Rendering frameworks such as \PixiJS provide developers with a COR to specify how objects should be displayed in terms of properties such as position, size, opacity, and tint~\cite{macklon2022automatically}.
The HTML5 \canvas is particularly useful for developing visual-centric web applications such as video games, data visualizations, and animations~\cite{macklon2023taxonomy, applecanvas}.

There are two types of rendered graphics in HTML5 \canvas applications:
\begin{enumerate}
    \item \assetbasedgraphics Graphics: To use \assetbasedgraphics graphics, the developer must first create or procure image assets for the display of objects such as Graphical User Interface (GUI) elements or game objects. Objects are then rendered by applying image transformations to image assets, after which they are displayed on the \canvas bitmap~\cite{mozillacanvasapitutorial, mozillawebglapitutorial}.
    \item \proceduralgraphics Graphics: Instead of (or in addition to) preparing image assets for an HTML5 \canvas application, a developer may choose to utilize \proceduralgraphics graphics.
    For \proceduralgraphics graphics, the display of objects on the \canvas is determined by algorithms that describe how to programatically ``draw'' objects that are then displayed on the \canvas~\cite{mozillacanvasapitutorial, mozillawebglapitutorial}.
\end{enumerate}

\subsection{Visual Bugs}
Visual bugs are mismatches between the expected and actual visual output in the graphics of an application~\cite{issa2012visual}.
In our prior work, we investigated testable \canvas issues and discovered four types of visual bugs in HTML5 \canvas applications: \State, \Rendering, \Layout, and \Appearance~\cite{macklon2023taxonomy}.
\State bugs are instances in which objects on the \canvas are displayed in the incorrect state, such as an object being invisible when it should be visible.
\Rendering bugs are instances in which an object contains artifacts or is distorted (e.g., blurry).
\Layout bugs are instances in which the positioning, sizing, or layering of an object is incorrect.
\Appearance bugs are instances in which the color, opacity, or tint of an object is incorrect.
Figure~\ref{fig:visualbugsamples} in Section~\ref{sec:methodology} shows an example of each of these four visual bug types.

\subsection{Detecting Visual Bugs in HTML5 \canvas Applications}~\label{subsec:detectingvisualbugsbackground}
Due to the visual-centric nature of \canvas applications, visual bugs are important to detect because such bugs can render a \canvas application useless.
It is challenging to automatically detect visual bugs in \canvas applications because they do not operate like traditional web applications.
Traditional web applications have contents that are represented in the DOM but the contents of HTML5 \canvas applications are not represented in the DOM.
HTML5 \canvas applications require different testing approaches because generic web testing approaches typically rely on analysis of the DOM and therefore do not work for the \canvas~\cite{macklon2023taxonomy}.
Snapshot testing is an approach that does not necessarily rely on the DOM, however direct screenshot-to-screenshot comparisons require the maintenance of visual test oracles which is tedious and time-consuming~\cite{bajammal2018web, leotta2014visual}.
Instead, the testing of \canvas applications is mostly done manually~\cite{macklon2023taxonomy}, and prior studies on \canvas testing have mostly focused on how to improve these manual processes in an industrial setting~\cite{viggiato2023prioritizing, viggiato2022identifying, viggiato2022using}.

\section{Related Work} \label{sec:related}
In this section, we describe prior work related to our study. 
We primarily discuss studies regarding HTML5 \canvas testing and studies that leverage VLMs for bug detection.
We also discuss related studies on automated web or GUI testing because HTML5 \canvas bugs share some similarities with generic web and GUI bugs~\cite{macklon2023taxonomy}.

\subsection{HTML5 \canvas Testing}\label{subsec:relatedworkcanvastesting}
Only two prior studies~\cite{bajammal2018web, macklon2022automatically} have investigated approaches for automatically testing the HTML5 \canvas.

Bajammal and Mesbah~\cite{bajammal2018web} propose an approach for automated testing of the \canvas by inferring the contents of the \canvas through traditional computer vision techniques, and report high accuracy in detecting \State visual bugs.
The experiments detailed in their study evaluate their approach with a single type of visual bug (\State)~\cite{macklon2023taxonomy} in five HTML5 \canvas applications; any other type of visual bug on the \canvas would interfere with their visual inference algorithms.
In contrast, our approach using VLMs can detect four types of visual bugs (\State, \Rendering, \Layout, and \Appearance)~\cite{macklon2023taxonomy} while maintaining high bug-free accuracy when evaluated on screenshots from a set of 20 HTML5 \canvas applications.

In our prior work, we propose an approach for automatically detecting visual bugs in HTML5 \canvas applications that use \assetbasedgraphics graphics~\cite{macklon2022automatically}, and report high accuracy when evaluating our approach across four types of visual bugs in a single toy \canvas game.
However, our prior approach requires that the \canvas application has image assets available to be leveraged as a visual test oracle, meaning it cannot be used to detect visual bugs in objects rendered using \proceduralgraphics graphics in HTML5 \canvas applications.
Additionally, our prior approach requires the creation and maintenance of complex image processing algorithms at the core of the testing approach.
The approach studied in this paper instead targets both \assetbasedgraphics and \proceduralgraphics graphics in \canvas applications and is much simpler to use than our prior approach because we leverage pre-trained VLMs instead of maintaining complex image processing algorithms.

\subsection{Leveraging VLMs for Software Testing}
Prior work has investigated leveraging VLMs to automatically detect software bugs without specifying a test oracle.

Taesiri et al.~\cite{taesiri2024glitchbench} leverage VLMs to perform automated question-answering (Q\&A) with screenshots from open-world video games to evaluate the accuracy of using various VLMs to detect open-world video game glitches.
They find that \gptfouro vastly outperforms all other studied VLMs for automatically describing video game glitches in open-world video games.
Our study uses VLMs in a similar way to their study but instead of benchmarking various VLMs on the task of describing video game glitches in open-world games, we focus on detecting visual bugs in HTML5 \canvas applications by leveraging a state-of-the-art VLM through prompting strategies that utilize readily-available context.

In another study, Taesiri and Bezemer~\cite{taesiri2024videogamebunny} train custom VLMs using various fine-tuning approaches, then perform automated Q\&A to describe video game screenshots in a similar fashion to the prior work by Taesiri et al.~\cite{taesiri2024glitchbench}.
They report improvements in accuracy for VLM video game understanding after fine-tuning an 8-billion parameter VLM on various video game-specific instruction-following datasets.
Our study instead focuses on detecting visual bugs in HTML5 \canvas applications, and we avoid the costs and deep learning expertise associated with training new models by leveraging a state-of-the-art pre-trained VLM.

Ju et al.~\cite{ju2024study} investigate leveraging VLMs to extract and analyze event sequences from screenshots (along with the screenshot itself) to detect non-crash functional bugs in mobile GUI applications.
They find that their approach can detect some non-crash functional bugs at the cost of many false positives.
However, requiring developers to review many false positives undermines the purpose of test automation~\cite{macklon2022automatically, alegroth2016maintenance}.
Our study instead targets visual bugs in HTML5 \canvas applications which are different from bugs in mobile GUI applications (though they may share some similarities)~\cite{macklon2023taxonomy}, and our experiments show that our approach for detecting visual bugs in \canvas applications can yield very high \Precision.

\subsection{Automated Web and GUI Testing}~\label{subsec:webtestingrelatedwork}
While HTML5 \canvas applications require different testing approaches from generic web applications, \canvas bugs share some similarities with web and GUI bugs~\cite{macklon2023taxonomy}.
Many studies have investigated automated web~\cite{dougan2014web, ricca2021web} and automated GUI testing~\cite{banerjee2013graphical} including many computer-vision based techniques~\cite{bajammal2020survey} for web and GUI testing.
In this section we compare and contrast our study with works that automatically generate test cases and works that perform similar forms of visual analysis in web and GUI applications.

Prior works have studied detecting visual bugs in web or GUI applications through various forms of snapshot testing~\cite{lin2014accuracy, hori2015oracle, mahajan2014finding, moran2018detecting, tanno2018support}.
Approaches for detecting visual bugs that leverage snapshot testing are prone to requiring oracle re-verification due to (often small) functionally correct differences in the graphics of the application that cause false positives to be reported by such approaches~\cite{bajammal2018web, leotta2014visual}.
The laborious process of continuously re-verifying snapshot testing oracles for HTML5 \canvas applications defeats the purpose of test automation~\cite{macklon2022automatically, alegroth2016maintenance}.
Our work avoids the issue of oracle re-verification by instead leveraging VLMs and readily-available information about the correct functionality of the HTML5 \canvas application to automatically detect visual bugs.

Prior works have studied automatically writing test cases for web and GUI applications or crawling web and GUI application states through a variety of methods.
These methods include: approaches for automatically constructing and exploring a state-flow graph of web and GUI applications~\cite{marchetto2008state, mesbah2008crawling, sun2008framework, mesbah2009invariant, biagiola2019diversity}, approaches for automatically generating test inputs for web and GUI applications~\cite{artzi2011framework, biagiola2017search, deiner2020search}, template-matching driven testing for GUI applications~\cite{chang2010gui, mozgovoy2018unity, yeh2009sikuli}, Region-based Convolutional Neural Network (R-CNN) driven testing for mobile GUI applications~\cite{andelkovic2019bait}, and LLM-driven~\cite{feng2024enabling} or VLM-driven~\cite{liu2024vision, shahbandeh2024naviqate} approaches for testing web and mobile GUI applications.
These prior works are complementary to our work on detecting visual bugs in HTML5 \canvas applications because automatically writing test cases or exploring application states could help developers apply the approach studied in our paper to a larger number of screenshots to automatically detect more visual bugs in HTML5 \canvas applications.

\section{Leveraging VLMs to Detect Visual Bugs}\label{sec:leveragingvlms}
VLMs are deep learning models that can process image and text inputs simultaneously, and in this paper we are interested in pre-trained VLMs such as \gptfouro.
Pre-trained VLMs offer a way to avoid explicitly specifying a visual test oracle by connecting text and vision inputs through learned embeddings of the concepts represented in those inputs~\cite{zhang2024vision}, enabling the output of a text-based visual analysis.
Given a screenshot of an application and a (text and/or image-based) description of the correct functionality, a VLM should be able to ``see'' the contents of the image(s) and connect those contents with the information provided in the text(s) to determine if a visual bug is present in the screenshot.
In our study, we leverage VLMs to detect visual bugs by emulating a user of a chat application such as OpenAI's ChatGPT.
We emulate a user of such a chat application by using the API offered by OpenAI to generate chat responses using pre-trained VLMs such as \gptfouro, enabling automated analysis of screenshots collected from \canvas applications.

\subsection{Prompting Strategies}
To address the possibility of distribution drift between images in the pre-training dataset of a VLM and the visual content of \canvas applications (see Section~\ref{sec:introduction}), we propose prompting strategies that provide additional inputs (\emph{context}).
Each prompting strategy utilizes different combinations of text and image data to help describe the application's expected functionality in addition to the screenshot for analysis.
In particular, we supply combinations of the README from the GitHub repository, the bug descriptions from a taxonomy of \canvas bugs~\cite{macklon2023taxonomy}, a \bugfree screenshot, and image assets.
We use the README and bug descriptions as text inputs, and we use the \bugfree screenshot and image assets as image inputs.
Table~\ref{tab:promptstrategies} shows each of our prompting strategies and which pieces of context were included for each strategy.
Our full prompts for each prompting strategy are provided in the appendix.

\begin{table}[!t]
\centering
\caption{Each of our prompting strategies shown with which pieces of context are included.}
\begin{tabular*}{\columnwidth}{@{} p{0.3\columnwidth}  @{\extracolsep{\fill}} p{0.12\columnwidth} @{} p{0.12\columnwidth} @{} p{0.12\columnwidth} @{} p{0.12\columnwidth} @{}}
\toprule
\textbf{Prompting strategy} & \textbf{README} & \textbf{Bug descriptions} & \textbf{Bug-free screenshot} & \textbf{Image assets} \\
\midrule
\promptzero &  &  &  &  \\
\promptone  & \multicolumn{1}{r}{\checkmark} &  &  &  \\
\prompttwo  & \multicolumn{1}{r}{\checkmark} & \multicolumn{1}{r}{\checkmark} &  &  \\
\promptthree & \multicolumn{1}{r}{\checkmark} & \multicolumn{1}{r}{\checkmark} & \multicolumn{1}{r}{\checkmark} & \\
\promptfour & \multicolumn{1}{r}{\checkmark} & \multicolumn{1}{r}{\checkmark} & \multicolumn{1}{r}{\checkmark} & \multicolumn{1}{r}{\checkmark} \\
\bottomrule
\end{tabular*}
\label{tab:promptstrategies}
\end{table}

\section{Constructing Our Evaluation Dataset} \label{sec:methodology}
\begin{figure*}[!t]
\centering
	\centering
	\includegraphics[width=\linewidth]{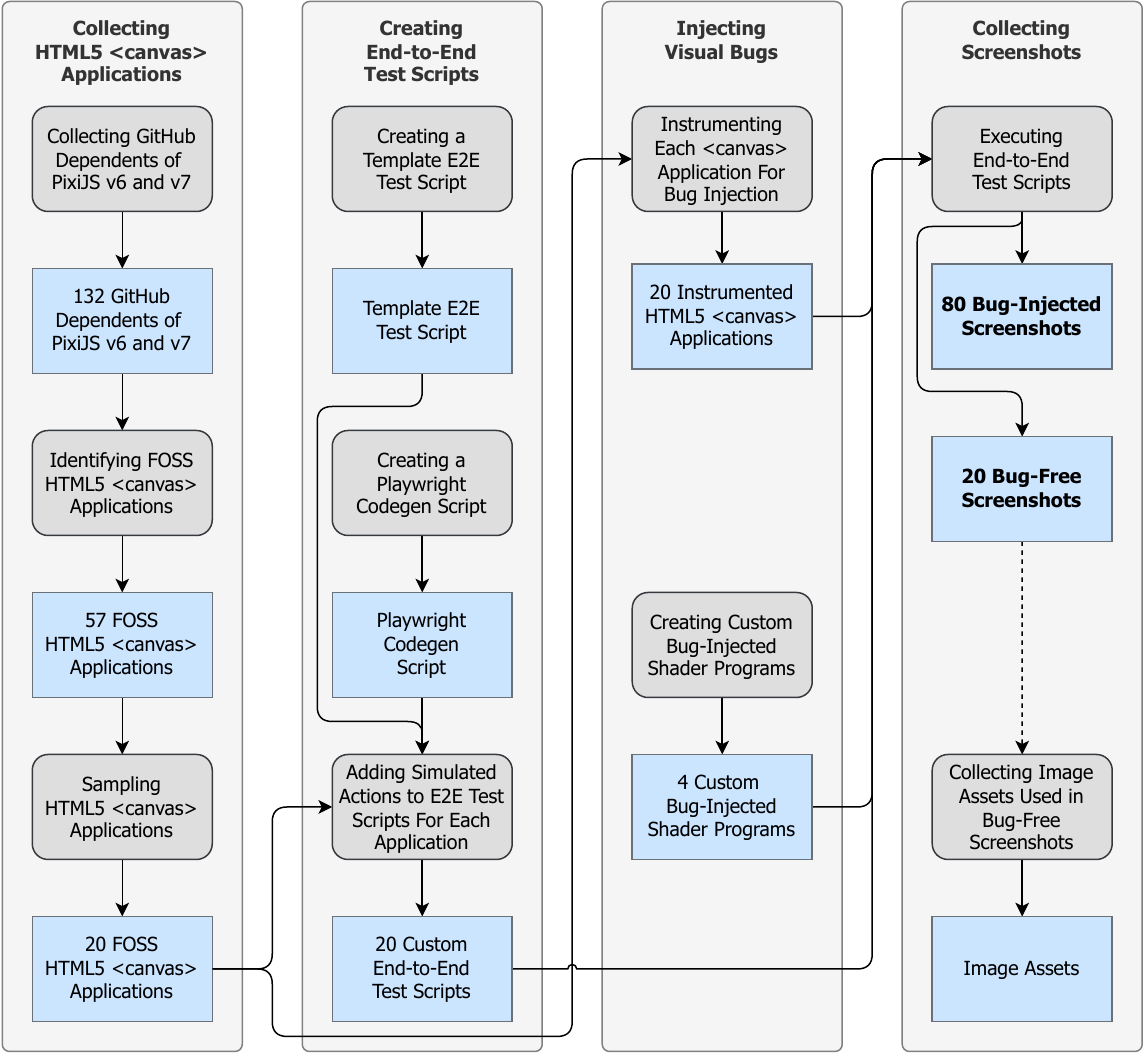}
	\caption{Overview of our dataset construction.}
	\label{fig:datasetconstruction}
\end{figure*}

In this section, we describe the approach we followed to construct a dataset of 20 \bugfree and 80 \buginjected screenshots from 20 FOSS HTML5 \canvas applications.
Figure~\ref{fig:datasetconstruction} shows an overview of our dataset construction.

\subsection{Collecting HTML5 \canvas Applications}
First, we collected and filtered GitHub repositories to produce a set of 57 FOSS HTML5 \canvas applications that use \PixiJS, a widely-used HTML5 \canvas rendering framework~\cite{macklon2022automatically}.
Then, for practical purposes, we sampled 20 FOSS HTML5 \canvas applications to collect screenshots from.

\subsubsection{Collecting GitHub Dependents of \PixiJS}
We began our process for collecting FOSS HTML5 \canvas applications by collecting GitHub repositories that are dependents of \PixiJS.
Because the \PixiJS rendering framework has undergone several major architectural changes since its inception, we chose to focus on the most recent versions of \PixiJS (\textsc{v6} and \textsc{v7}).
Both \PixiJS \textsc{v6} and \textsc{v7} are similar technically and allow visual bugs to be injected in the same way using our custom framework, described in Section~\ref{subsec:injectingbugs}.
By focusing on the two most recent (major) versions of \PixiJS, we also ensure that our study captures modern \canvas applications.

\paragraph{Crawling the GitHub Dependency Graph}
We collected GitHub repositories that are dependents of \PixiJS by crawling the GitHub Dependency Graph using the \textsc{github-dependents-info}\footnote{\url{https://bit.ly/github-dependents-info}} Python library.
The GitHub Dependency Graph can be used to identify which GitHub repositories depend on (i.e., use) a specified GitHub repository~\cite{githubdependencygraph}.
We crawled the list of dependents of \PixiJS on GitHub in July 2023, producing a list of 16,189 GitHub repositories.

\paragraph{Automatically Filtering on Stars and \PixiJS Package}
We set a threshold of 11 stars for our set of 16,189 GitHub repositories that use \PixiJS to remove toy repositories~\cite{munaiah2017curating, maj2024fault, kalliamvakou2014promises}.
The number of GitHub stars that a repository has is a metric that is widely-used to represent the popularity of repositories~\cite{munaiah2017curating, maj2024fault}, and popular repositories are more mature compared to the entire population of repositories~\cite{maj2024fault}.
A threshold of 11 stars simultaneously enabled the removal of many toy repositories while keeping the largest reasonable set of applications that were to be manually analyzed.
We also filtered our set of collected GitHub repositories based on the specific \PixiJS package(s) depended upon by the project, as it is possible to optionally require only portions of the \PixiJS framework.
We kept GitHub dependents of \PixiJS that use the \texttt{pixi.js} or \texttt{pixi.js-legacy} packages as these packages are used for making \PixiJS applications, as opposed to leveraging part of the \PixiJS framework's functionality.
After filtering GitHub repositories, we were left with 682 GitHub repositories that are dependents of \PixiJS and have at least 11 GitHub stars.

\paragraph{Identifying the \PixiJS Version Used}
We used the GitHub API to identify which \PixiJS version each of the 682 collected GitHub repositories used.
We searched for mentions of \textit{`pixi'} in the \texttt{package.json}, \texttt{package-lock.json}, and \texttt{yarn.lock} files of each repository.
We then filtered the text matches from the search results to only include the texts that included one of previously specified \PixiJS packages (\textit{`pixi.js'} or \textit{`pixi.js-legacy'}) as a substring.
We then manually verified the filtered text matches for each repository and automatically extracted the version numbers from the text matches.
Using this method, we were able to automatically identify \PixiJS versions in 358 of the 682 repositories.

\paragraph{Automatically Filtering on \PixiJS Version Used}
For the 358 GitHub repositories that we were able to automatically identify a \PixiJS version, we automatically filtered any repository that did not match the \PixiJS versions specified for our study.
In particular, we only kept GitHub repositories that matched a \PixiJS version like \textit{`6.*.*'} or \textit{`7.*.*'}.
After filtering for \PixiJS \texttt{v6} or \texttt{v7}, we were left with 88 repositories.

\paragraph{Manually Filtering on \PixiJS Version Used}
For the remaining 324 repositories without an automatically identifiable \PixiJS version, we manually inspected the code and documentation contained in each repository on GitHub to determine the \PixiJS version used.
Of these 324 repositories, 44 used \PixiJS \texttt{v6} or \texttt{v7}.
While many of the remaining repositories used a different version of \PixiJS, some did not specify any \PixiJS version in their GitHub repository, i.e., the code did not contain any usage of \PixiJS.
The lack of \PixiJS version for some repositories could be due to repositories dropping support for \PixiJS without the GitHub Dependency Graph reflecting this change.
Combined with the repositories identified through automatic filtering, we had a total of 132 GitHub repositories.

\subsubsection{Identifying FOSS HTML5 \canvas Applications}
With 132 of our collected GitHub repositories verified as using \PixiJS \texttt{v6} or \texttt{v7}, we proceeded to the next stage of manual filtering in which we confirmed whether each repository contained a FOSS HTML5 \canvas application.
Examples of repositories that were filtered out at this stage include desktop \textsc{NodeJS} applications that require \texttt{node-canvas} or game engine frameworks built on top of \PixiJS.
We manually filtered our set of 132 repositories according to the following selection criteria:

\begin{itemize}
    \item The repository does not contain proprietary assets.
    \item The repository contains a \canvas application.
    \item The \canvas application does not use external (paid) APIs.
    \item The \canvas application uses \PixiJS for its graphics.
    \item The \canvas application does not already contain (visual) bugs.
\end{itemize}

After this manual filtering, we were left with 57 FOSS HTML5 \canvas applications.

\subsubsection{Sampling FOSS HTML5 \canvas Applications}
For practical purposes, we manually performed stratified sampling to select 20 \canvas applications based on their graphics type (\proceduralgraphics or \assetbasedgraphics graphics) and the type of application (e.g., game or data visualization).
Table~\ref{tab:targetsvsactuals} shows the proportions of each graphics type and application type across the set of 57 applications paired with the proportions across the sample of 20 \canvas applications.
Table~\ref{tab:subsetcanvasapplications} shows a brief description of each of the 20 FOSS HTML5 \canvas applications in our dataset.
With a set of 20 FOSS HTML5 \canvas applications, we proceeded to collect \bugfree and \buginjected screenshots of the applications.

\begin{table}[!t]
\centering
\caption{Proportion of FOSS HTML5 \canvas applications belonging to each application type and graphics type for all 57 applications and our sample of 20 applications.}
\begin{tabular*}{\columnwidth}{@{} l l @{\extracolsep{\fill}} p{0.26\linewidth} p{0.28\linewidth} @{}}
\toprule
\textbf{Type of} & \textbf{Category} & \textbf{Proportion in all 57 apps (\%)} & \textbf{Proportion in our sample of 20 apps (\%)} \\
\midrule
\multirow{4}{*}{\textbf{Application}} & \animations & \multicolumn{1}{r}{14} & \multicolumn{1}{r}{20} \\
& \datavisualization & \multicolumn{1}{r}{22} & \multicolumn{1}{r}{20} \\
& \visualeditor & \multicolumn{1}{r}{24} & \multicolumn{1}{r}{25} \\
& \game & \multicolumn{1}{r}{40} & \multicolumn{1}{r}{35} \\
\midrule
\multirow{2}{*}{\textbf{Graphics}} & \assetbasedgraphics & \multicolumn{1}{r}{55} & \multicolumn{1}{r}{55} \\
& \proceduralgraphics & \multicolumn{1}{r}{45} & \multicolumn{1}{r}{45} \\ \bottomrule
\end{tabular*}
\label{tab:targetsvsactuals}
\end{table}

\begin{table}[!t]
\centering
\caption{Summary of our sample of 20 FOSS HTML5 \canvas applications.}
\begin{tabular*}{\columnwidth}{@{} l @{\extracolsep{\fill}} l @{}}
\toprule
\textbf{GitHub repository} & \textbf{Application description} \\
\midrule
\textsc{Aidymouse/Hexfriend} & Hexmap editor \\
\textsc{aldy-san/zero-neko} & Kanji typing game \\
\textsc{chase-manning/react-photo-studio} & Photoshop clone \\
\textsc{coderetreat/coderetreat.org} & Conway's Game of Life \\
\textsc{dimforge/rapier.js} & Physics simulations \\
\textsc{equinor/esv-intersection} & Well intersection visualization \\
\textsc{getkey/ble} & Platformer level editor \\
\textsc{ha-shine/wasm-tetris} & Tetris clone \\
\textsc{higlass/higlass} & Genome data visualization \\
\textsc{mehanix/arcada} & Architectural blueprints editor \\
\textsc{MichaelMakesGames/reflector} & Base-builder game \\
\textsc{ourcade/ecs-dependency-injection} & Breakthrough clone \\
\textsc{p5aholic/playground} & Particle simulations \\
\textsc{PrefectHQ/graphs} & Graph diagrams \\
\textsc{solaris-games/solaris} & Idle strategy game \\
\textsc{starwards/starwards} & Space map editor \& simulations \\
\textsc{tulustul/ants-sandbox} & Ant colony simulations \\
\textsc{uia4w/uia-wafermap} & Wafer map visualization \\
\textsc{VoiceSpaceUnder5/VoiceSpace} & Virtual world \\
\textsc{Zikoat/infinite-minesweeper} & Minesweeper clone \\
\bottomrule
\end{tabular*}
\label{tab:subsetcanvasapplications}
\end{table}

\subsection{Creating End-to-End Test Scripts}
We created a custom end-to-end (E2E) test script for each HTML5 \canvas application to collect \bugfree and \buginjected screenshots.
E2E test scripts automatically launch and navigate a \canvas application with simulated user actions to enable testing of the functionality experienced by the end-user.
To create E2E test scripts for each of the 20 HTML5 \canvas applications, we used the \textsc{Playwright} testing framework (v1.44.1) for \textsc{NodeJS} (v20.11.1).
In this section we explain how we leveraged \textsc{Playwright} to create custom E2E test scripts for HTML5 \canvas applications.

\subsubsection{Creating a Template E2E Test Script}
Before creating a custom E2E test script for each application, we first created a template E2E test script.
This template E2E test script contains code common to all custom E2E test scripts, such as accepting command line arguments, launching the browser, and opening the \canvas application.
Additionally, this template E2E test script includes the code required to use our custom visual bug injection framework (see Section~\ref{subsec:injectingbugs}), such as importing dependencies and a custom function call to take a screenshot with (or without) a visual bug injected.

\subsubsection{Creating a \textsc{Playwright} \textsc{Codegen} Script}
We leveraged the \textsc{Codegen}\footnote{\url{https://playwright.dev/docs/codegen}} features of \textsc{Playwright} to record some user actions in each \canvas application.
While \textsc{Playwright} relies largely on the DOM to simulate user actions in web applications, there is some functionality offered for recording mouse clicks (including \texttt{x,y} coordinates) on the \canvas.
To use the \textsc{Codegen} features of \textsc{Playwright}, we created a custom script that opened each \canvas application and then paused the execution to allow the recording of mouse clicks.

\subsubsection{Adding Simulated Actions to E2E Test Scripts For Each Application}
For each \canvas application, we modified a copy of the template E2E test script to include simulated user actions.
We added simulated user actions using code leveraging the \textsc{Playwright} API through a combination of: (1)~recording mouse clicks using \textsc{Playwright Codegen}, and (2)~by manually writing code for actions such as mouse scrolls and key presses. 
To use our \textsc{Playwright Codegen} script for recording mouse clicks, we first had to set up and serve the \canvas applications. 

\paragraph{Running and Serving Each Application}
We ran and served each \canvas application to enable the use of \textsc{Playwright Codegen}.
The process described here was also used to run and serve each \canvas application while executing our custom E2E test scripts as described in Section~\ref{subsec:executingtestscripts}.
We first set-up the environments to execute each application locally by following steps detailed in the documentation available in each \canvas application's GitHub repository.
For example, some applications required the configuration of a web server and a database server.
In some cases, the steps detailed were insufficient for running the application and most often the solution was to install and use a specific version of \textsc{NodeJS} for installing dependencies and running the application's web server (many of the \canvas applications used \textsc{NodeJS} for the web server).
We only ran and served a single \canvas application at any given time.

\paragraph{Simulating User Actions Using Playwright}
While serving each \canvas application, we iteratively updated a copy of our custom \textsc{Codegen} script as we created our custom E2E test scripts in order to pause the execution at different points and record additional mouse clicks.
After recording mouse clicks in a \canvas application, we copied the code generated using \textsc{Playwright Codegen} into each \canvas application's respective custom E2E test script.
These recorded mouse clicks combined with the manual addition of other simulated user inputs (such as mouse scrolling and key presses) provided E2E test scripts that could automatically navigate through each \canvas application to ensure that the sampled visual states were representative of each \canvas application.
In some cases, the application by default did not display many objects within the \canvas because it is intended to start from a blank state (e.g., hexmap editor without any objects yet placed onto the hex tiles).
For such applications, we added some additional simulated user actions to add some content to the \canvas and ensure the test case was representative of a real-world end-to-end test case.

\subsection{Injecting Visual Bugs}\label{subsec:injectingbugs}
We created a custom framework for performing Just-in-Time (JiT) visual bug injection in HTML5 \canvas applications that are built with \PixiJS \texttt{v6} or \texttt{v7}.
In this section we describe how we instrumented each \canvas application with custom code to enable our custom visual bug injection framework and how we created custom \buginjected shaders.

\subsubsection{Instrumenting Each \canvas Application For Bug Injection}
We instrumented each \canvas application such that a JavaScript reference to the \PixiJS \texttt{PIXI} object or the \PixiJS \texttt{PIXI.Application} instance was available in the global window object of the web browser when running the \canvas application as required for our custom visual bug injection framework.
Developers of \canvas applications that use the \PixiJS framework will commonly import the \PixiJS framework into a \canvas application's code as the \texttt{PIXI} object, but in some cases we found that it was simpler to copy a reference to the \canvas application's \texttt{PIXI.Application} instance to the global window object because of how the \canvas application was bundled for execution.
A reference to either the \texttt{PIXI} object or \texttt{PIXI.Application} instance in the global scope of the \canvas application is sufficient for our custom visual bug injection framework to function for that application.

\subsubsection{Creating Custom Bug-Injected Shaders}\label{subsubsec:creatingcustomshaders}
During the execution of an E2E test script utilizing our custom framework, the custom framework is loaded into the client-side context and looks for a reference to the \texttt{PIXI} object or \texttt{PIXI.Application} instance in the global window object to enable overrides of rendering functions provided by \PixiJS.
In particular, when performing JiT visual bug injection, our framework overrides the WebGL graphics shaders used by \PixiJS with our custom \buginjected shaders.
These shaders include both a fragment shader and vertex shader that are written in Open\underline{GL} \underline{S}hading \underline{L}anguage (GLSL) and compiled by WebGL in the browser.
By overriding the shader program used by \PixiJS when rendering a frame, we ensure that we do not inject any bugs into the COR, and only inject bugs into the bitmap of the \canvas.
Table~\ref{tab:visualbugtypesandinjection} shows the four visual bug types described in our previously published taxonomy of testable HTML5 \canvas issues~\cite{macklon2023taxonomy} mapped to how we injected a visual bug of that type.
Figure~\ref{fig:visualbugsamples} shows screenshots of one instance of each of the four visual bug types across various \canvas applications.

\begin{table*}[t]
\centering
\caption{Descriptions of our custom WebGL shaders used to inject visual bugs in HTML5 \canvas applications.}
\begin{tabular*}{\textwidth}{@{} l @{\extracolsep{\fill}} p{0.8\linewidth} @{}}
\toprule
\textbf{Bug type}  & \textbf{Description of our change(s) to default \PixiJS WebGL shaders} \\
\midrule
\State & Turn visible objects invisible by multiplying the \texttt{vColor} by a very small value (0.001) in the vertex shader so that the opacity is effectively zero. \\
\midrule
\Rendering & Introduce artifacts into objects by multiplying the \texttt{gl\_FragColor} for specific bands using a scaling factor in the fragment shader. \\
\midrule
\Layout & Scale objects to a small size and incorrect position by introducing a scaling factor for \texttt{gl\_Position} in the vertex shader. \\
\midrule
\Appearance & Add additional transparency into objects by multiplying the \texttt{vColor} by 0.5 in the vertex shader, halving the opacity. \\
\bottomrule
\end{tabular*}
\label{tab:visualbugtypesandinjection}
\end{table*}

\subsection{Collecting Screenshots}
Having instrumented 20 HTML5 \canvas applications and created 20 custom E2E test scripts plus four buggy shaders, we proceeded to collect \bugfree and \buginjected screenshots from each \canvas application.
Our screenshots of the \canvas bitmap are each paired with their respective COR.
We stored the ground truth labels of each screenshot (e.g., \Rendering bug) in the filename of each screenshot.
Additionally, for \buginjected screenshots, we dynamically insert a property named \texttt{\_\_injected\_by\_pixi\_visual\_bugger\_\_} (set to \texttt{true}) on objects in the COR that are injected with visual bugs on the \canvas.

\subsubsection{Executing End-to-End Test Scripts}\label{subsec:executingtestscripts}
We executed our E2E test scripts in a Chromium (build 125.0.6422.26) browser window with dimensions of 720 pixels vertically and 1280 pixels horizontally, similar to our prior work~\cite{macklon2022automatically}. 
We ran each E2E test script in five different ways: once without any visual bugs injected, and then again to separately inject each of the four visual bug instances corresponding to the four studied types of visual bugs (\Layout, \Rendering, \Appearance, and \State).

\paragraph{Executing End-to-End Test Scripts Normally}
We executed each E2E test script once normally without any visual bugs injected.
Then, we manually analyzed each of the 20 \bugfree screenshots to ensure that no visual bugs were present in the screenshots.

\paragraph{Executing E2E Test Scripts With Visual Bugs Injected}\label{subsubsec:executingwithvisualbugs}
For each of our four custom buggy shaders, we executed each E2E test script with that buggy shader program to produce four \buginjected screenshots per \canvas application for a total of 80 \buginjected screenshots.
We manually analyzed each of the 80 \buginjected screenshots and verified that the injected visual bugs were visible in the screenshot and could be detected by a human tester.

\begin{figure*}[!tp]
\centering
\begin{subfigure}[t]{0.4875\linewidth}
    \centering
    \includegraphics[width=\textwidth]{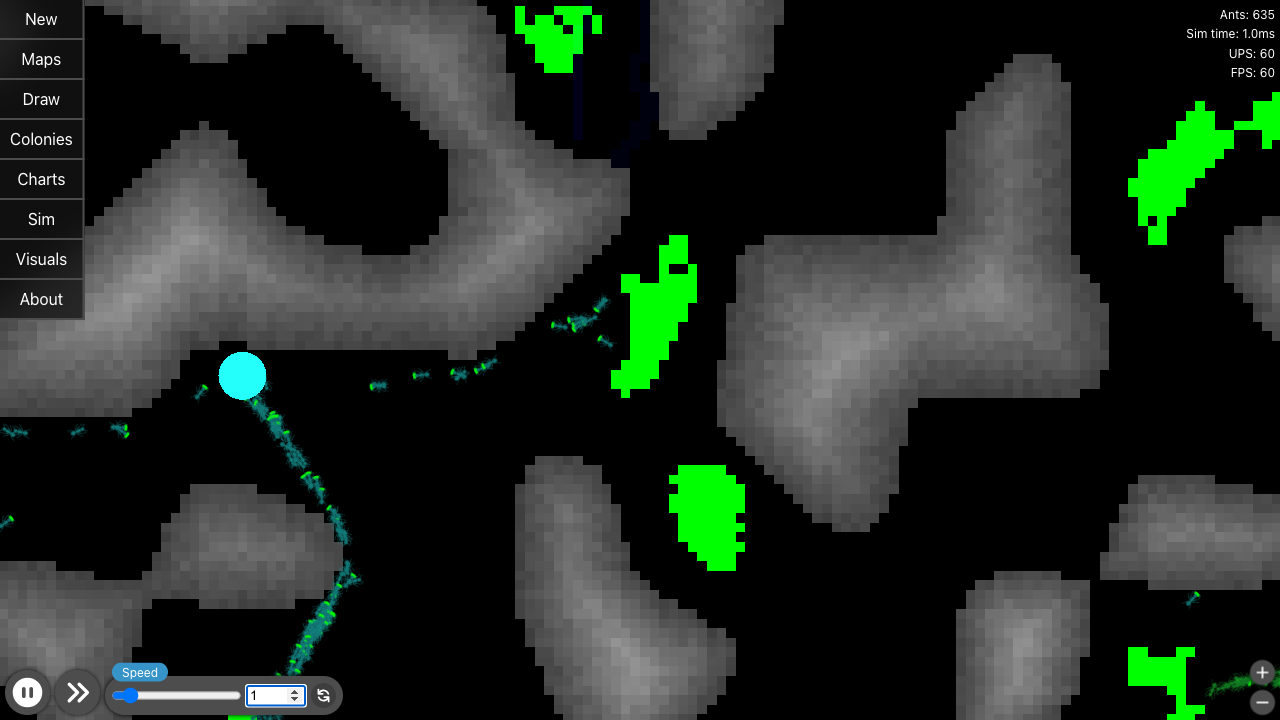}
    \caption{Screenshot of \textsc{tulustul/ants-sandbox} without visual bugs.}
\end{subfigure}
\hfill
\begin{subfigure}[t]{0.4875\linewidth}
    \centering
    \includegraphics[width=\textwidth]{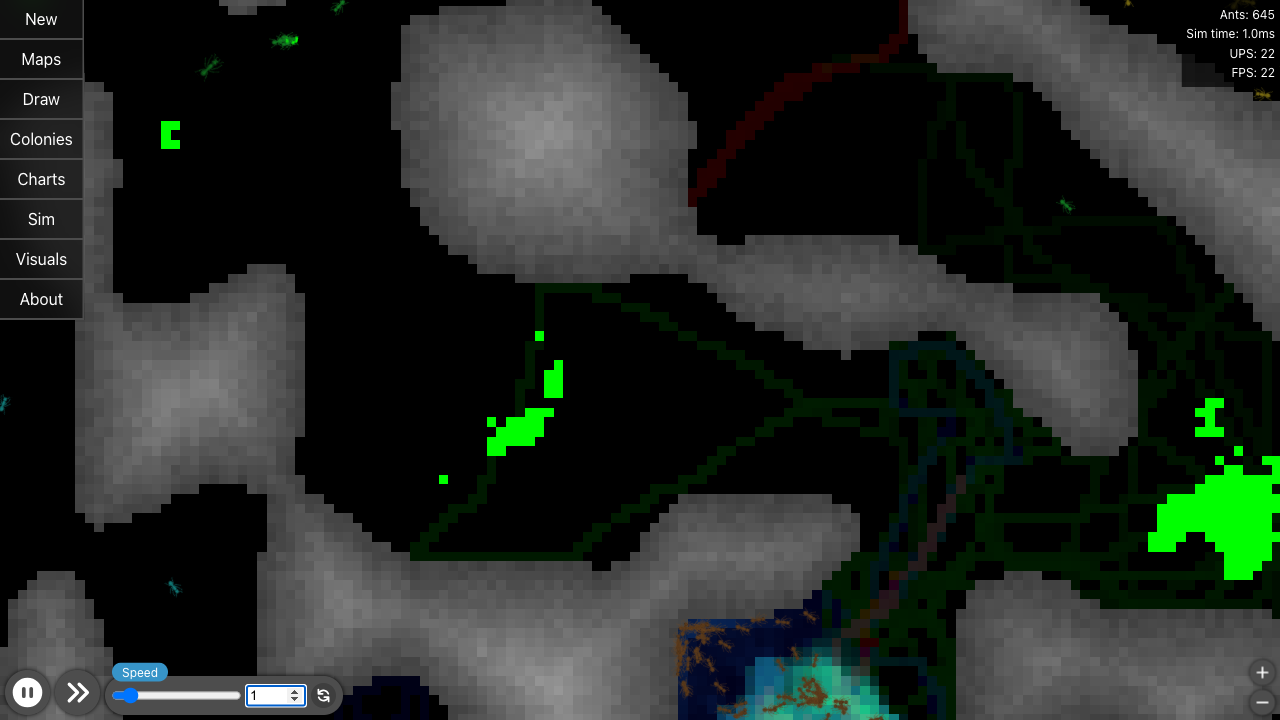}
    \caption{\State bug injected into \textsc{tulustul/ants-sandbox}. The ant colony nodes (circles) are missing when they should be displayed.}
\end{subfigure}

\begin{subfigure}[t]{0.4875\linewidth}
    \centering
    \includegraphics[width=\textwidth]{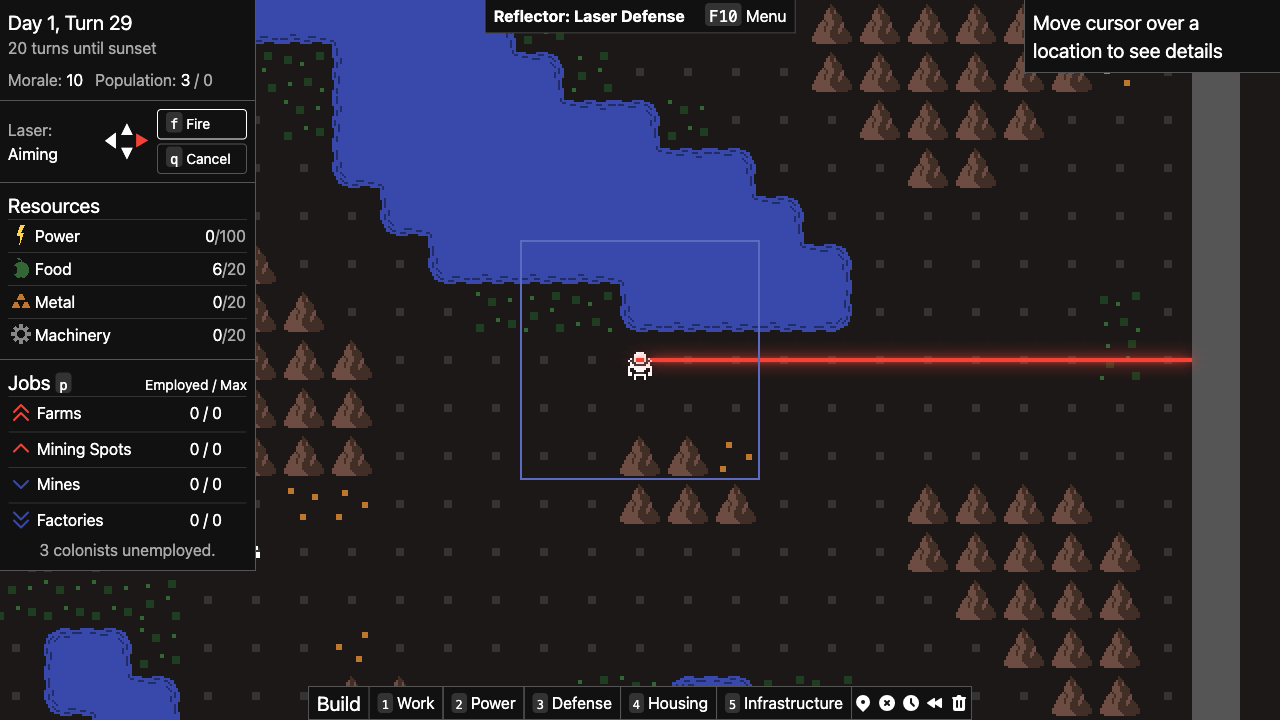}
    \caption{Bug-free screenshot of \textsc{MichaelMakesGames/reflector}.}
\end{subfigure}
\hfill
\begin{subfigure}[t]{0.4875\linewidth}
    \centering
    \includegraphics[width=\textwidth]{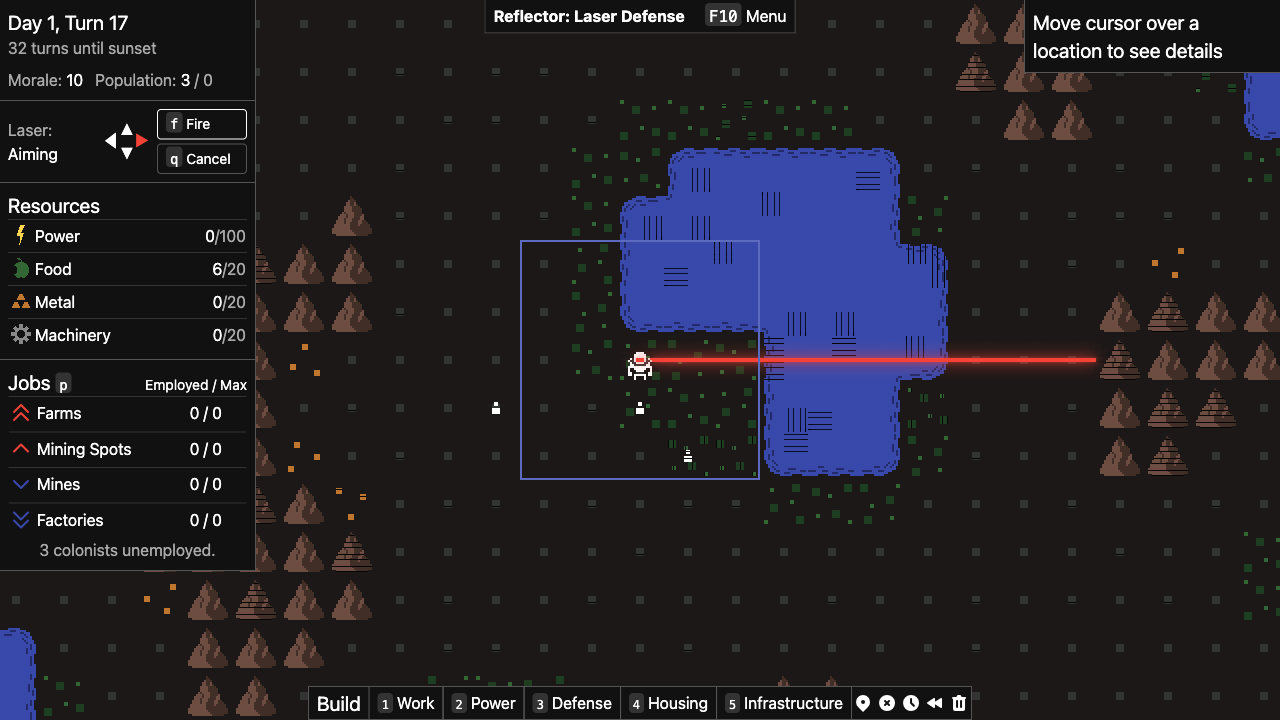}
    \caption{\Rendering bug injected into \textsc{MichaelMakesGames/reflector}. Tiles contain artifacts (black lines).}
\end{subfigure}

\begin{subfigure}[t]{0.4875\linewidth}
    \centering
    \includegraphics[width=\textwidth]{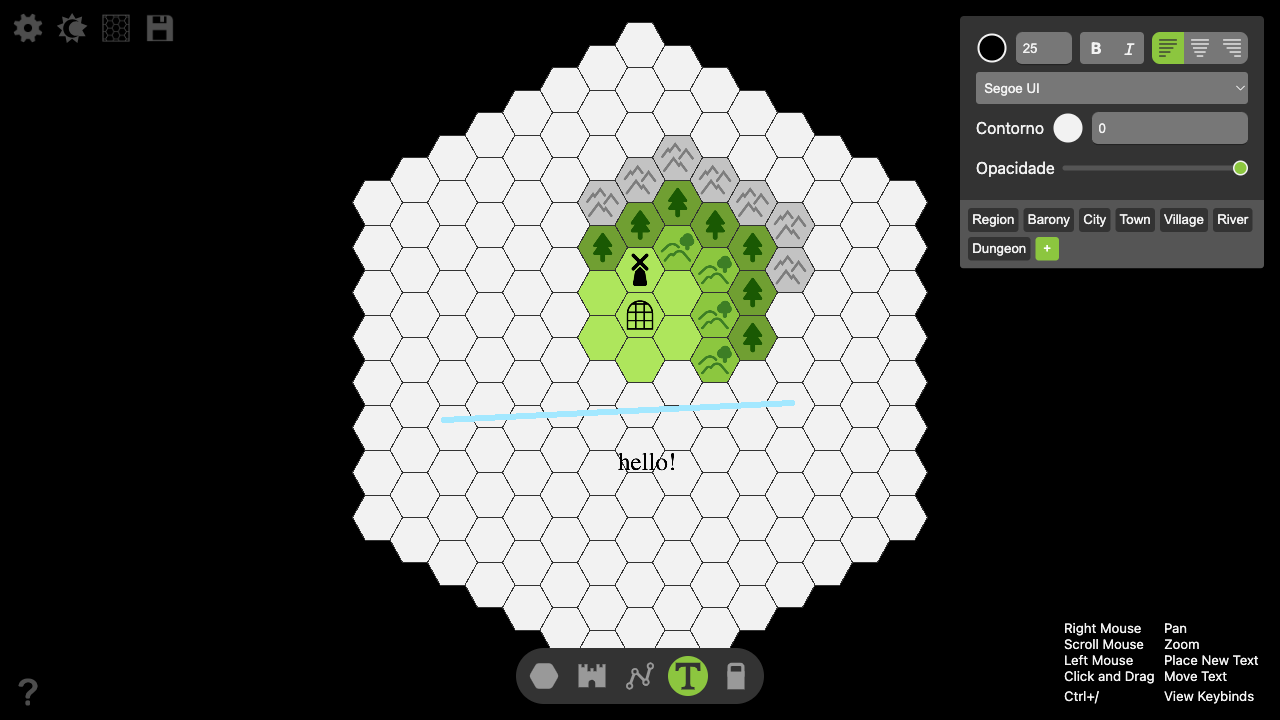}
    \caption{Bug-free screenshot of \textsc{Aidymouse/Hexfriend}.}
\end{subfigure}
\hfill
\begin{subfigure}[t]{0.4875\linewidth}
    \centering
    \includegraphics[width=\textwidth]{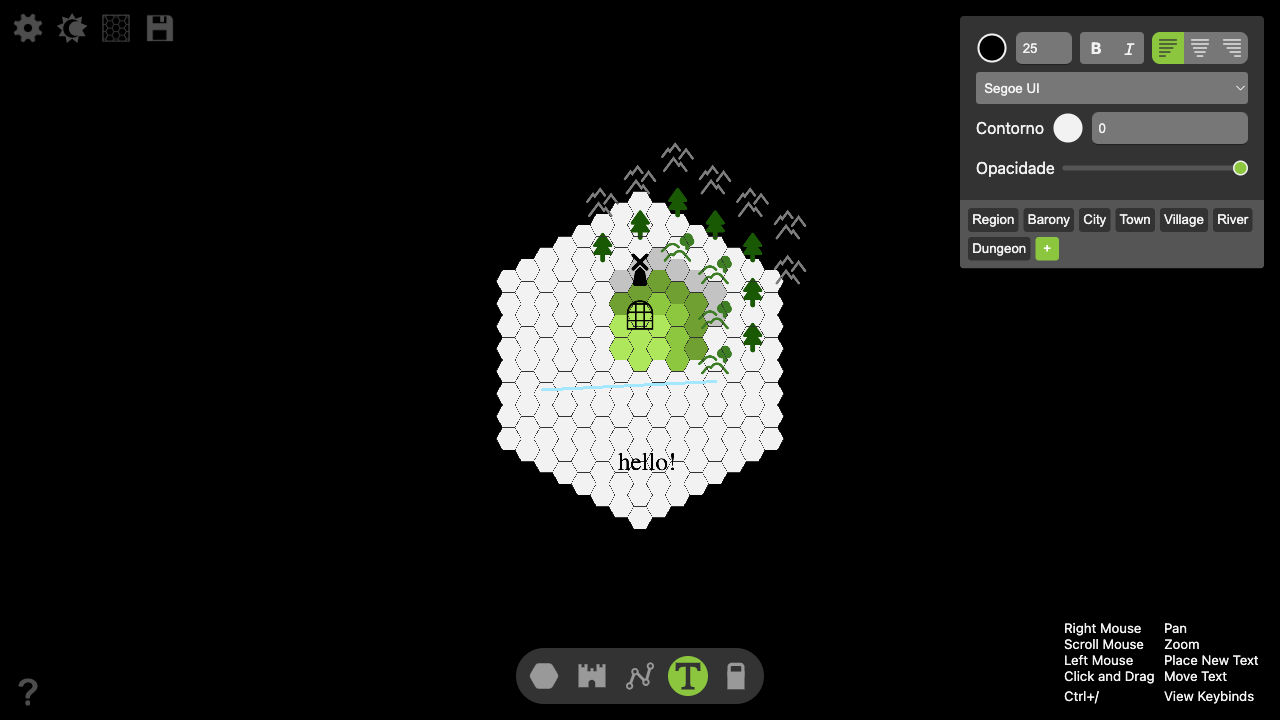}
    \caption{\Layout bug injected into \textsc{Aidymouse/Hexfriend}. Hexagonal tiles are in an incorrect position and are incorrectly sized.}
\end{subfigure}

\begin{subfigure}[t]{0.4875\linewidth}
    \centering
    \includegraphics[width=\textwidth]{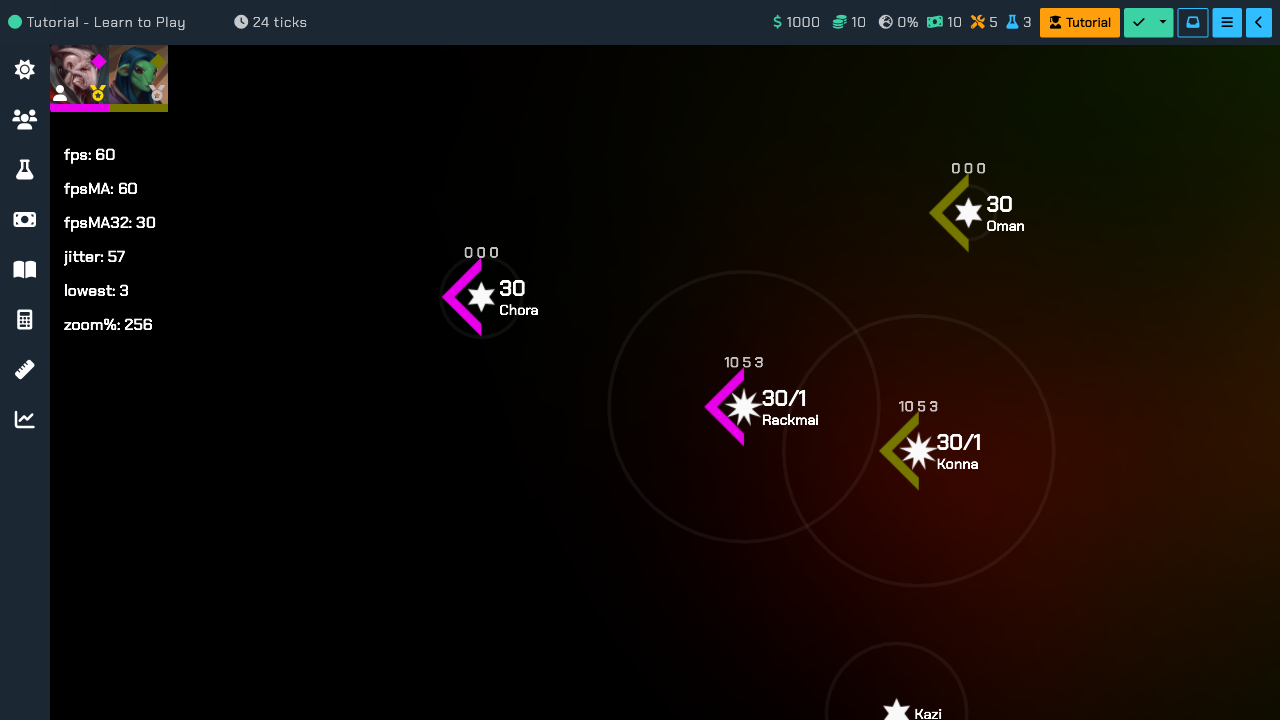}
    \caption{Bug-free screenshot of \textsc{Solaris-Games/Solaris}.}
\end{subfigure}
\hfill
\begin{subfigure}[t]{0.4875\linewidth}
    \centering
    \includegraphics[width=\textwidth]{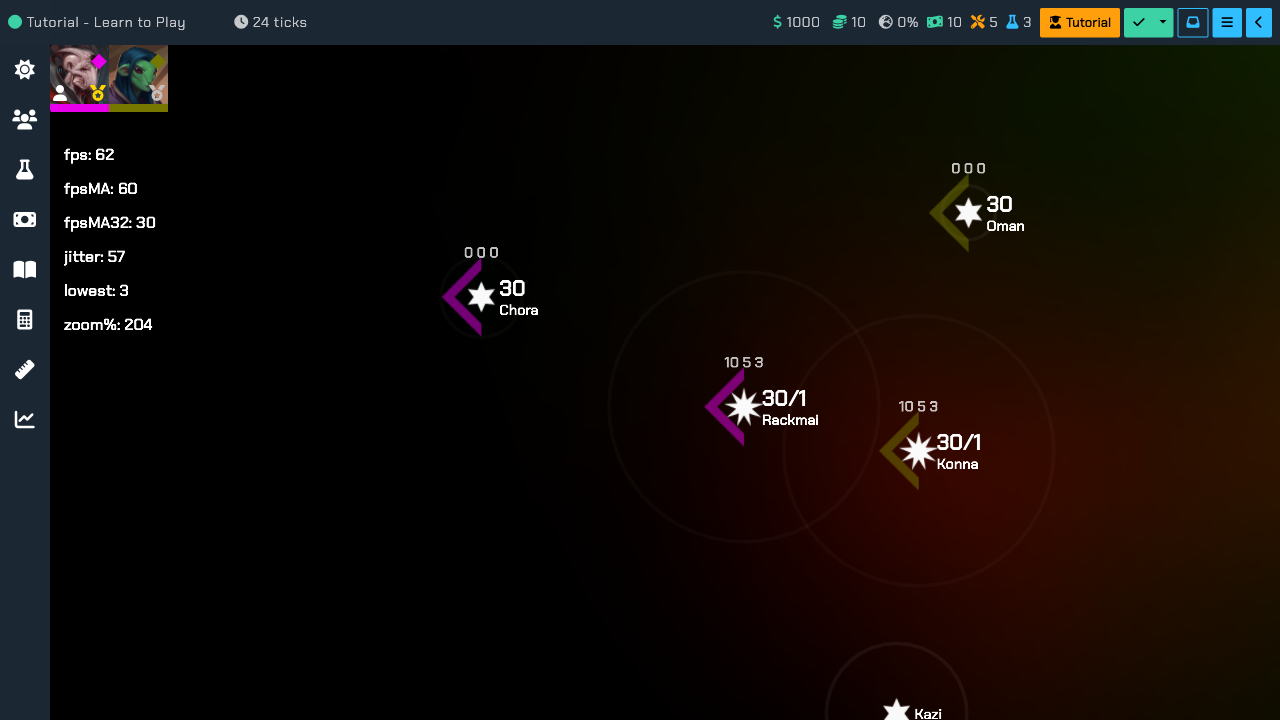}
    \caption{\Appearance bug injected into \textsc{Solaris-Games/Solaris}. The pink and yellow faction indicators are partially transparent.}
\end{subfigure}
\caption{Four bug-free screenshots ((a), (c), (e), (g)) and four bug-injected screenshots ((b), (d), (f), (h)) of \canvas applications collected using our custom framework. Each bug-injected screenshot is paired with a description of the injected visual bug.}
\label{fig:visualbugsamples}
\end{figure*}

\subsubsection{Collecting Image Assets Used in Bug-Free Screenshots}
In addition to collecting screenshots (and their respective CORs), we also collected the image assets used during each normal (i.e., \bugfree) E2E test execution for the 11 \canvas applications that use \assetbasedgraphics graphics.
We collected the image assets to run experiments with the image assets included as context (see Section~\ref{sec:experiments}).

\paragraph{Determining Which Image Assets Were Used}
We parsed each COR from each test execution into a table saved in \texttt{CSV} format using a custom script.
We then manually reviewed each parsed COR to determine which image assets were used during each test case execution for the \canvas application with \assetbasedgraphics graphics.

\paragraph{Rasterising the Image Assets}
Then, for each image asset that was used, we manually collected a bitmap copy of that image asset, so that it could be used as input for the VLMs.
For image assets that were already bitmaps (such as those in Portable Network Graphics (PNG) format), we simply copied the original image asset.
For image assets that were vector graphics (e.g., Scalable Vector Graphics (SVG) format) we first rasterised the image asset by opening it inside of the preview pane of Google Drive while using the Google Chrome browser and then exported a bitmap in PNG format by taking a screenshot (using macOS's built-in screenshot functionality) of the rasterised image asset.

\section{Experimental Set-Up} \label{sec:experiments}
To evaluate the effectiveness of leveraging VLMs for automatically detecting visual bugs in HTML5 \canvas applications, we prompted a state-of-the-art VLM, \gptfouro, using various prompting strategies while supplying each screenshot from our dataset of 100 screenshots.
We ran the experiments with the first four of our prompting strategies (\promptzero, \promptone, \textit{README+ BugDescriptions}, and \promptthree) across all 100 screenshots.
However, the prompting strategy that used image assets (\promptfour) was only used for the 11 \canvas applications that had image assets, across a total of 55 screenshots.
The range of lengths of READMEs supplied per application ranged from 130 tokens to 2420 tokens (when tokenized with the \texttt{gpt-4o-2024-08-06} tokenizer).
We calculated the number of README tokens using the \textsc{tiktoken} Python library.
We used the OpenAI API to prompt the \gptfouro model snapshot \texttt{gpt-4o-2024-08-06} on October 6th, 2024.

\subsection{Running the Experiments}
For each of the 20 \bugfree and each of the 80 \buginjected screenshots in our dataset, we sequentially generated two separate chat completions (i.e., text completions generated using a VLM that has been finetuned for chat applications) per prompting strategy to automatically detect visual bugs.
First, we generated a chat completion using \gptfouro that contained a visual analysis based on the input screenshot image plus additional context as specified for each prompting strategy in Table~\ref{tab:promptstrategies}.
Then, we used the structured outputs mode offered through the OpenAI API to generate a second chat completion using \gptfouro in a separate chat thread that converted the visual analysis into a consistent structure in JavasScript Object Notation (JSON) format.
Extracted answers in JSON format had the following two fields: 
\begin{itemize}
    \item \texttt{bool\_did\_detect\_visual\_bug}: indicates whether a visual bug was described in the text generated using \gptfouro.
    \item \texttt{string\_description\_of\_visual\_bug}: contains the text description of the visual bug generated using \gptfouro.
\end{itemize}
During answer extraction, our prompts included instructions to leave the text description of the bug empty if no visual bug was described in the prior chat completion. 
Our full prompt for answer extraction is provided in the appendix.
We ran our experiments with the model temperature set to the default value (one) to emulate a chat conversation using \gptfouro through a ChatGPT user interface.
We completed four repetitions of our experiments to help account for variability in VLM output in our results.
The monetary cost of running our experiments using the OpenAI API was less than \$4.50 United States Dollars (USD) per repetition.

\subsubsection{Running the Experiments to Understand the Impact of Different Pieces of README Text}
To understand the impact of different pieces of text in the README files on the capabilites of VLMs to detect visual bugs, we performed an ablation study with prompting strategy \promptone.
First, we ran our main experiments (as described in the previous paragraph) and determined which \canvas applications had the three highest differences and three lowest differences in recall when comparing \promptone to \promptzero.
The recall of these six applications was the most affected by including the README file(s) in the prompt to the VLM.
We then proceeded to manually split each application's README file(s) into two versions for two new prompting strategies: \promptablationA and \promptablationB.
We included text we hypothesized was useful for visual bug detection (such as descriptions of application functionality) in \promptablationA and included text we hypothesized was not useful (such as descriptions of how to run the application) in \promptablationB.
Table~\ref{tab:ablationreadmetokens} shows the number of tokens in (i.e., lengths of) each customized version of the README file(s) for each application in our ablation study.
We then ran our experiments with these prompting strategies in the same way as described in the previous paragraph.

\begin{table*}
\centering
\caption{Lengths of customized README file(s) per-application used for \promptablationA and \promptablationB in the ablation study experiments.}
\begin{tabular*}{\linewidth}{@{} l r @{\extracolsep{\fill}} r @{}}
\toprule
 & \multicolumn{2}{l}{\textbf{Number of tokens}} \\ 
\textbf{Application} & \textbf{\promptablationA} & \textbf{\promptablationB} \\
\midrule
\textsc{chase-manning-react-photo-studio} &	 15    &	364   \\
\textsc{ha-shine-wasm-tetris}           &	11     &	85    \\
\textsc{higlass-higlass}                &	134    &	1888  \\
\textsc{p5aholic-playground}            &	0      &	194   \\
\textsc{starwards-starwards}            &	114    &	1254  \\
\textsc{VoiceSpaceUnder5-VoiceSpace}    &	115    &	2323  \\
\bottomrule
\end{tabular*}
\label{tab:ablationreadmetokens}
\end{table*}

\subsection{Evaluating the VLM Outputs}
To evaluate the outputs generated using \gptfouro in our experiments, we first compared the truth value in the output field \texttt{bool\_did\_detect\_visual\_bug} to the ground truth labels that were available in our dataset of 100 screenshots.
The ground truth labels in our dataset describe whether a screenshot contains a visual bug, and if so, the ground truth label also describes the type of visual bug that was injected into that screenshot.
For predictions that a visual bug was present in the screenshot, we manually reviewed the description of the visual bug in the text of the output field \texttt{string\_description\_of\_visual\_bug} to compare it to the screenshot (plus the ground truth label).
If the output text generated with the VLM contained an accurate description of the visual bug that we observed in the screenshot, then that prediction was marked as correct (True Positive (TP)), and otherwise it was considered to be a False Positive (FP).
Other FPs were incorrect predictions that a visual bug was present in screenshots that were actually \bugfree, whereas True Negatives (TN) were correct predictions that a \bugfree screenshot did not contain a bug.
False Negatives (FN) were incorrect predictions that there was not a visual bug present in screenshots that did contain a visual bug.

\subsection{Evaluation Metrics}
For each prompting strategy, we measured accuracy, precision, recall, and pass@$k$.
Accuracy represents what proportion of samples were correctly classified (as containing or not containing a visual bug).
Recall represents what proportion of visual bugs were detected.
Precision represents the proportion of correct predictions that a visual bug is present out of all predictions that a visual bug is present including incorrect predictions, i.e., false positives.
Pass@$k$ represents whether any of the generated responses out of $k$ number of responses to the same input are correct.

A testing approach that only attains high recall (but low precision) would report many false positives and waste developer time as they diagnose the real visual bug (if one is present at all).
Meanwhile, a testing approach that only attains high precision (but very low recall) would not be very useful for detecting visual bugs.
We measured the accuracy for the \bugfree category of screenshots to help understand how often visual bugs were predicted to be present when in fact no visual bug was present in the screenshot.
For a testing approach to be useful, \bugfree accuracy must be very high to ensure that developers do not waste time dealing with many false positives in cases where no visual bug is present at all.
We measured pass@$k$ to understand how well leveraging VLMs to detect visual bugs works if generating multiple outputs for the same input, which could be an approach to handle the non-deterministic nature of VLM outputs.

We measured accuracy, precision and recall on a per-bug type basis across all 20 HTML5 \canvas applications for each of the four repetitions of experiments.
We also measured each prompting strategy's accuracy on a per-application basis across the bug-free category combined with the four visual bug type categories for each of the four repetitions of experiments.
We calculated overall accuracy per prompting strategy by computing the accuracy across all 100 screenshots with all four repetitions.
We calculated pass@$k$ for each screenshot using the formulation provided by Chen et al.~\cite{chen2021evaluating}, and we computed overall pass@$k$ by computing the arithmetic mean over the individual pass@$k$ values for all 100 screenshots.

Accuracy, precision, recall, and pass@$k$ are each calculated as follows:
\begin{align*}
\text{Accuracy}  = & \text{\space}\frac{\text{TN} + \text{TP}}{\text{TN} + \text{TP} + \text{FP} + \text{FN}} \\[4pt]
\text{Precision}  = & \text{\space}\frac{\text{TP}}{\text{TP} + \text{FP}} \\[4pt]
\text{Recall}  = & \text{\space}\frac{\text{TP}}{\text{TP} + \text{FN}} \\[4pt]
\text{Pass@$k$} = & \begin{cases}
1 & \text{for }(c > 0) \land ((n-c) < k) \text{,}\\[4pt]
1 - \displaystyle\prod_{i=n-c+1}^{n}{1 - \dfrac{k}{i}} & \text{for }(c > 0) \land ((n-c) \geq k)\text{,}\\[4pt]
0 & \text{for }c = 0\text{.}
\end{cases}
\end{align*}
where,
\begin{align*}
n  & \text{ is the number of responses generated (n=4 repetitions in our case),} \\[4pt]
c  & \text{ is the number of correct responses, and} \\[4pt]
k  & \text{ is the number of responses considered for correctness (i.e., $k$ in pass@$k$).}
\end{align*}

\section{Results} \label{sec:results}
\begin{figure}[!t]
\centering
\includegraphics[width=\columnwidth]{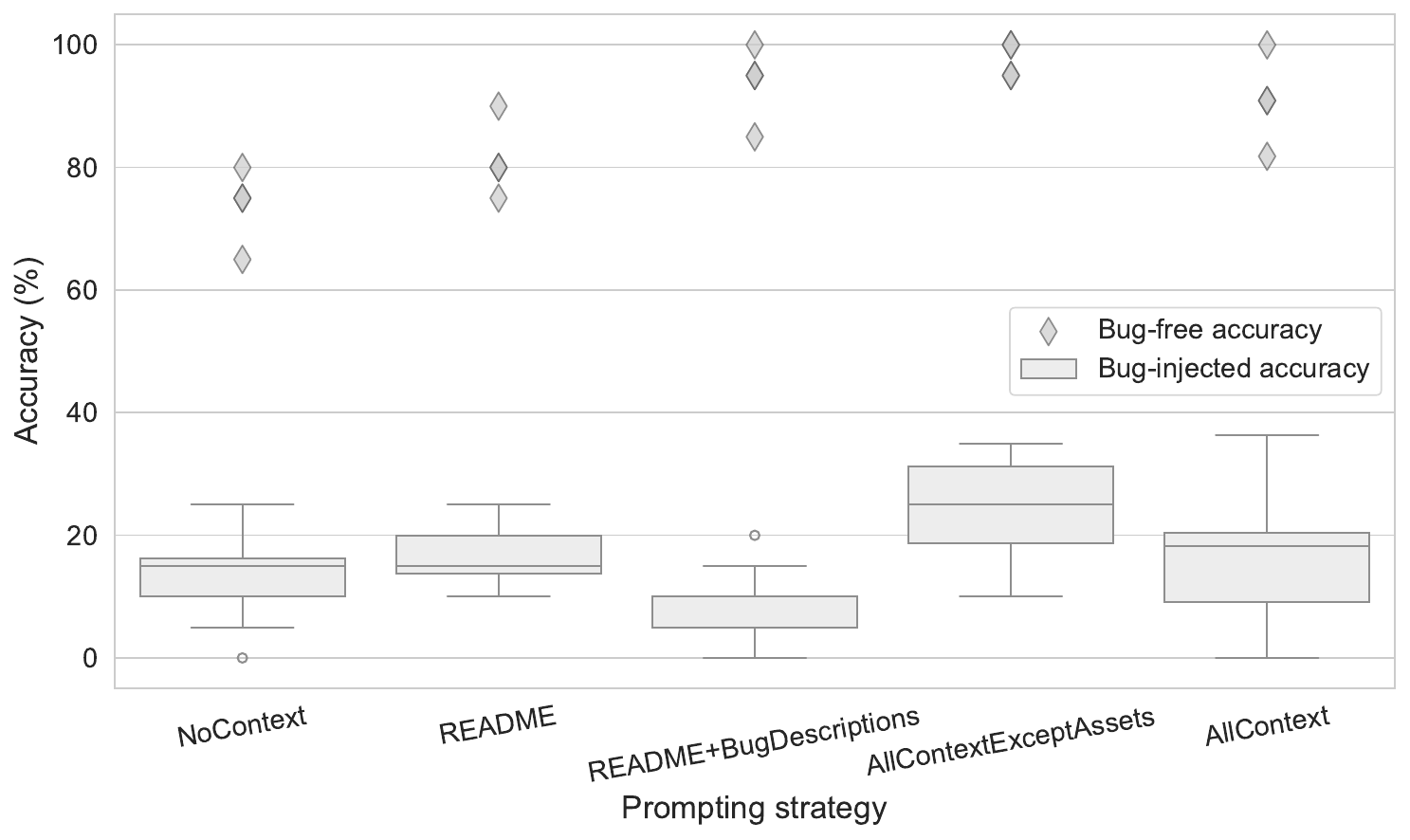}
\caption{Distributions of accuracy (\%) yielded in experiments using VLMs to detect visual bugs with various prompting strategies. Accuracy is computed over the set of 20 applications. The bug-free accuracies are shown as scatter plots with bug-injected accuracies shown as box plots on a shared set of axes. There are four bug-free accuracy values per prompting strategy. Each box plot represents a distribution of 16 values (four bug types multiplied by four repetitions).}
\label{fig:resultsA}
\end{figure}

\begin{figure}[!t]
\centering
\includegraphics[width=\columnwidth]{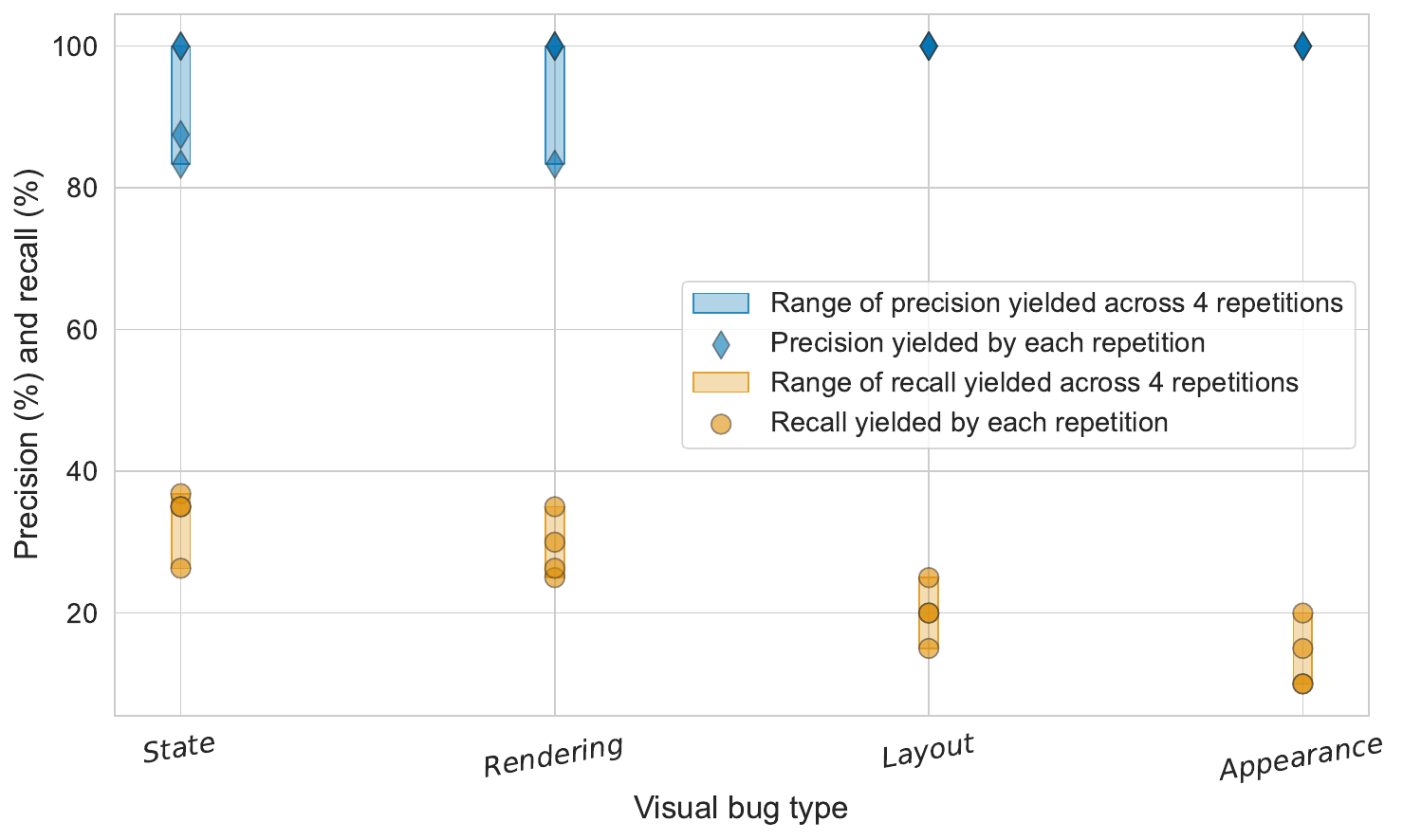}
\caption{Distributions of precision (\%) and recall (\%) per bug type when using VLMs to detect visual bugs with the prompting strategy \promptthree. Precision and recall are computed over the set of 20 applications. Precision and recall are represented on a shared set of axes as scatter plots with point clouds representing the ranges of values. There are four precision values and four recall values per bug type (from four repetitions of experiments).}
\label{fig:resultsB}
\end{figure}

In this section we discuss our results by revisiting and answering the research questions defined in Section~\ref{sec:introduction} of our paper.
We present our findings for each research question and suggest future research directions for each finding.
For each research question we also provide a summary of implications for developers of HTML5 \canvas applications.
Table~\ref{tab:resultsPassatk} shows the average pass@$k$ values paired with standard deviations across all 100 screenshots.
Figure~\ref{fig:resultsA} shows the distributions of accuracy, precision, and recall for each prompting strategy across four repetitions of experiments.
Figure~\ref{fig:resultsB} shows the distributions of precision and recall for each injected visual bug type when using the best-performing prompting strategy (\promptthree) across four repetitions of experiments.
Figure~\ref{fig:resultsC} shows the distributions of accuracy for each application across all four visual bug types when using the best-performing prompting strategy (\promptthree) across four repetitions of experiments.
The results of our ablation study are displayed in Figure~\ref{fig:resultsablation} which shows the differences in per-application recall for each of \promptablationA and \promptablationB when compared to the baseline prompting strategy \promptzero.

\begin{table*}[!t]
\centering
\caption{Pass@$k$ per-prompting strategy. The arithmetic mean of pass@$k$ values across the 100 screenshots is shown in column ``Avg.'', and the standard deviation of 100 pass@$k$ values is shown in column ``Std.''.}
\begin{tabular*}{\textwidth}{@{} l r @{\extracolsep{\fill}} r r @{\extracolsep{\fill}} r r @{\extracolsep{\fill}} r @{}}
\toprule
 & \multicolumn{2}{l}{\textbf{Pass@1}} & \multicolumn{2}{l}{\textbf{Pass@2}}  & \multicolumn{2}{l}{\textbf{Pass@4}} \\
\textbf{Prompting strategy} & \textbf{Avg.} & \textbf{Std.} & \textbf{Avg.} & \textbf{Std.} & \textbf{Avg.} & \textbf{Std.}\\
\midrule
\promptzero & 26 & 38 & 32 & 43 & 38 & 49 \\
\promptone & 29 & 40 & 35 & 44 & 42 & 50 \\
\prompttwo & 26 & 41 & 29 & 43 & 32 & 47 \\
\promptthree & 39 & 46 & 43 & 48 & 46 & 50 \\
\promptfour & 32 & 42 & 38 & 46 & 44 & 50 \\
\bottomrule
\end{tabular*}
\label{tab:resultsPassatk}
\end{table*}

\subsection*{RQ1: \rqone}

\textbf{VLMs can be leveraged to detect visual bugs in HTML5 \canvas applications by providing sufficient context.}
By utilizing a prompting strategy that provides sufficient context about the correct functionality of the \canvas application (including a \bugfree screenshot from the application), we find that VLMs can be leveraged as useful tools to detect visual bugs with relatively high accuracy.
The best overall performing prompting strategy, \promptthree, yields up to 100\% accuracy for the \canvas application \texttt{chase-manning/react-photo-studio}.
As can be seen in Figure~\ref{fig:resultsA}, it is also promising that \promptthree consistently yields high bug-free accuracy when aggregated across all 20 applications, ranging from 95\% to 100\%.
When aggregated across all 100 screenshots, \promptthree provides an overall accuracy of 39\%; an absolute increase in accuracy of 13\% when compared to the baseline prompting strategy \promptzero.
As Table~\ref{tab:resultsPassatk} shows, the large standard deviations in pass@$k$ values across the 100 screenshots indicates that there is a large amount of variability in the capabilities of VLMs to detect visual bugs.
We discuss this variability from a per-bug type and per-application basis in several of our following findings within this section of our paper.
Regardless, the overall (average) pass@$k$ values for \promptthree have absolute increases of 3\% to 4\% each time $k$ is doubled for our evaluated range of $k$ values (1, 2, 4).
This increase in pass@$k$ with increasing $k$ indicates that generating multiple outputs with a VLM per screenshot could be a possible approach for increasing bug detection rates with VLMs.

\emph{Future research directions:}
Our evaluation of VLMs for detecting visual bugs focuses on providing additional information as context in prompts to the VLM, rather than fine-tuning a VLM.
While providing application-specific information in the prompts to a VLM improves visual bug detection, it could be even more useful to fine-tune a VLM with application-specific data.
For example, a VLM could be fine-tuned using combinations of \bugfree screenshots and data available in the COR.
Future research should evaluate the benefits of fine-tuning a VLM for detecting visual bugs in \canvas applications, and compare the costs of utilizing such a fine-tuned model to increasing context during inference.

\textbf{\State visual bugs are the most successfully detected bug type when leveraging VLMs.}
Figure~\ref{fig:resultsB} shows the ranges and values of precision and recall attained per-bug type with the best overall performing prompting strategy (\promptthree).
This prompting strategy (\promptthree) provided an average of 33\% \Recall for screenshots containing \State visual bugs.
However, the average \Recall attained for the other bug types range from 14\% to 30\%.
Clearly, of the four studied visual bug types, \State bugs are the most easily detected bug type when leveraging VLMs.
However, the average precision of detecting \State bugs is lower than the other bug types. 
One possible reason that \State bugs are the easiest to detect may be that VLMs are better suited to detecting the visibility of objects than detecting differences in the appearance, layout, and rendering of those objects.
On the other hand, it could simply be that \State bugs generally cause the most obvious overall visual differences.

\emph{Future research directions:}
A deeper investigation into why \State bugs were the most easily detected bug type in our experiments may be useful in understanding how to improve visual bug detection leveraging VLMs.
Perhaps there are some parts of the context that better provide information on the correct functionality of the \canvas application in terms of \State visual bugs as opposed to the other bug types.
Understanding how individual types and instances of context impacted the resulting recall and precision values on a per bug type basis would be valuable when developing new prompting strategies.

\textbf{Using VLMs to detect \Rendering and \Layout visual bugs is promising.}
The best overall performing prompting strategy (\promptthree) provided 30\% average recall for \Rendering visual bugs and a 20\% average recall for \Layout visual bugs.
While these average recall values are not as high as those attained for \State visual bugs, they are still promising.

\emph{Future research directions:}
The detection of \Rendering bug instances may benefit from some kind of cropping or zooming augmentation before being supplied to a VLM for analysis, because such bugs might show up in the details of objects as can be seen in Figure~\ref{fig:visualbugsamples}b.
The detection of \Layout bug instances may instead benefit from some kind of visual annotation representing the expected bounding boxes (i.e., locations and sizes) of objects on the \canvas.
Therefore, future research could consider how to better target individual visual bug types on a type-by-type basis by applying data augmentations to screenshots before prompting the VLM.
Such an approach should ensure that they do not substantially decrease precision in the pursuit of higher recall for specific bug types.

\textbf{\Appearance visual bugs are the most challenging to detect with VLMs.}
\Appearance bugs were the most challenging to detect with an average recall of only 14\%.
However, detecting 14\% automatically is still better than the alternative of requiring all \Appearance bugs to be caught through manual testing.

\emph{Future research directions:}
Leveraging VLMs requires particular improvement for detecting \Appearance bugs.
Often, the appearance of an object on the \canvas is largely determined by combination of the values of the object's properties such as tint and opacity, which are available in the COR.
Providing such information as context to a VLM may prove useful in improving the detection of \Appearance visual bugs.

\begin{figure*}[!t]
\centering
\includegraphics[width=\linewidth]{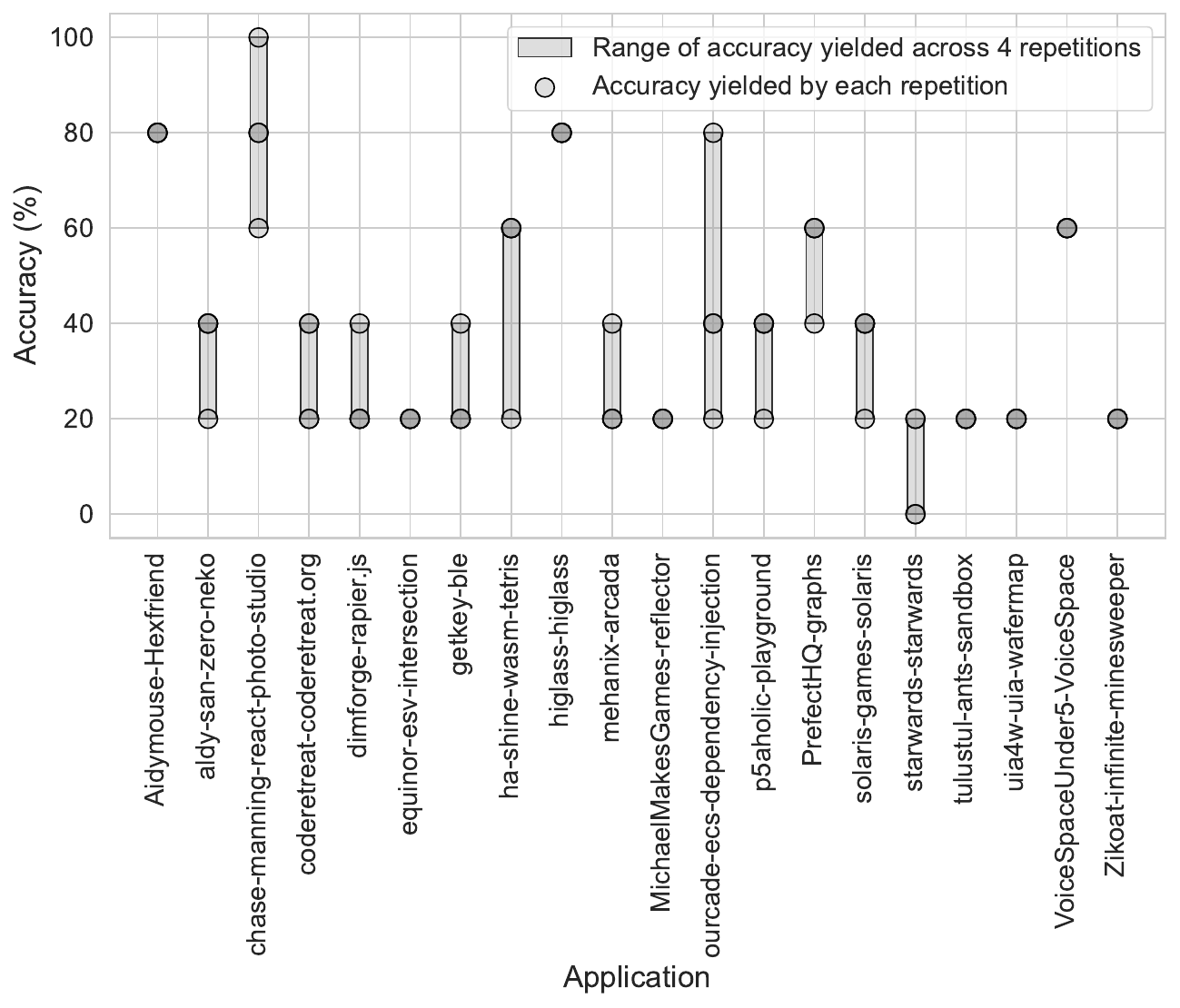}
\caption{Distributions of accuracy (\%) per-application when using VLMs to detect visual bugs with the prompting strategy \promptthree. Per-application accuracy is computed over the 100 screenshots (for each repetition) and accuracy values are represented as scatter plots with point clouds representing the range of values. There are four accuracy values per application (from four repetitions of experiments).}
\label{fig:resultsC}
\end{figure*}

\begin{figure*}[!t]
\centering
\includegraphics[width=\linewidth]{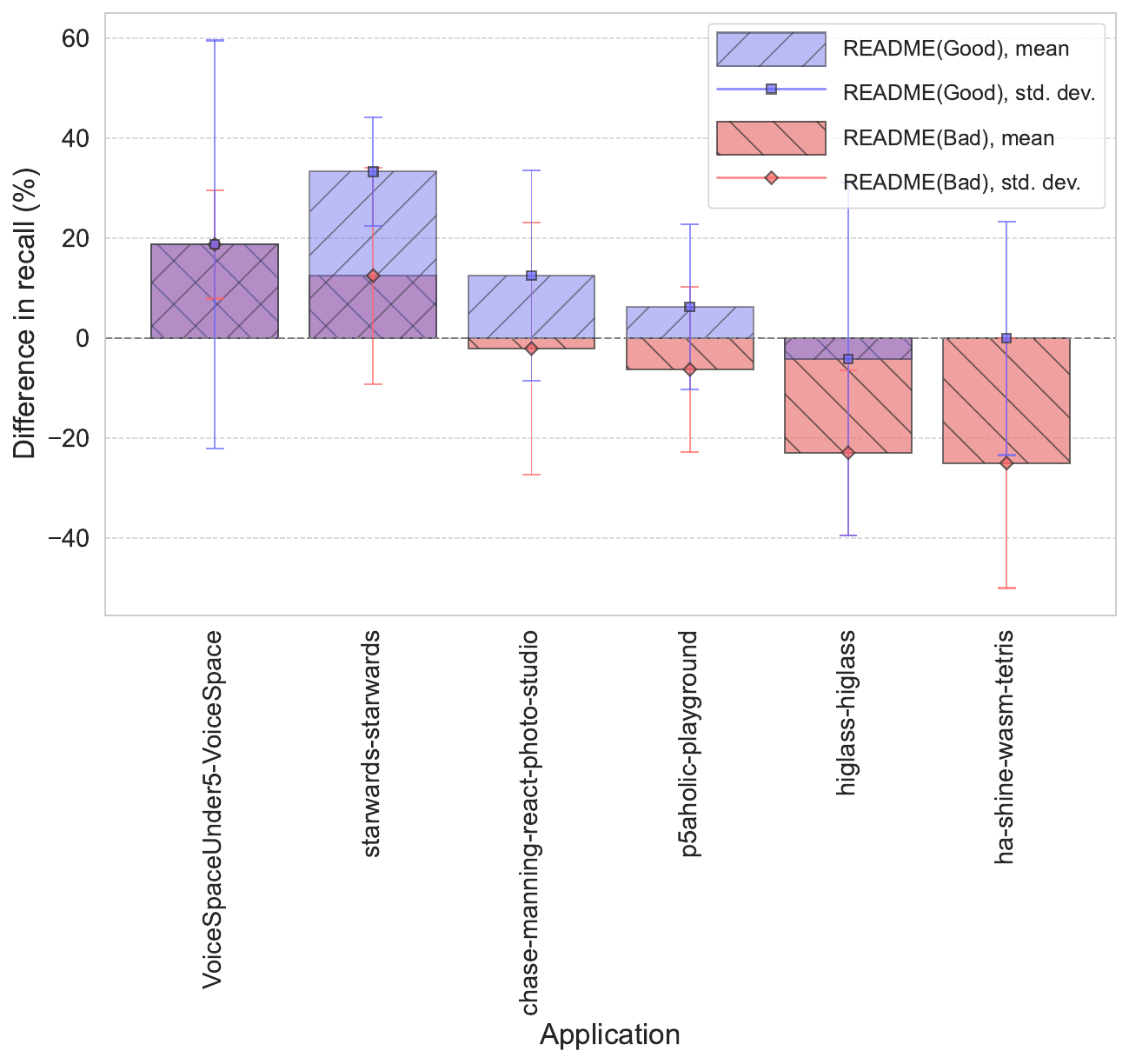}
\caption{Differences in recall (\%) per-application when using VLMs to detect visual bugs with the prompting strategies \promptablationA and \promptablationB each compared to \promptzero across 30 screenshots from 6 HTML5 \canvas applications. The average (mean) differences are represented with the bar plots, and the standard deviations (std. dev.) are represented with the error bars which are centered on a dot at the mean difference. Different hatching patterns and colours on the bar plots differentiate the two prompting strategies and where the strategies' differences overlap.}
\label{fig:resultsablation}
\end{figure*}

\textbf{The capabilities of VLMs to detect visual bugs varies on a per-application basis.}
As can be seen in Figure~\ref{fig:resultsC}, there is much variability in the visual bug detection results attained depending upon which application was being tested in our experiments.
While \promptthree can yield up to 100\% accuracy (and 100\% pass@1) for a \canvas application, accuracy was as low as 20\% on some repetitions for many of the applications.
This may be due to differences in visual styles across the 20 studied HTML5 \canvas applications.
However, the type of information in each application's README file may vary in terms of how informative it was to explain the correct functionality.
The results of our ablation study, shown in Figure~\ref{fig:resultsablation}, indicate that different parts of the application README file(s) can be more (or less) useful for prompting a VLM to detect visual bugs, and that the README has varying impact on recall depending upon the application.
While it is not exactly clear from our results which parts of the README file(s) are helpful, in most cases information about the correct functionality of the application was helpful (positive difference in recall), while information about how to set-up, install, and run the application was unhelpful (negative difference in recall).
Additionally, the amount of information in the README file(s) varies per application, as can be seen for several applications in Table~\ref{tab:ablationreadmetokens}.
Therefore, the variability in accuracy across applications is not only a result of differences in the applications themselves, but also a result of differences in the contextual information per application.

\emph{Future research directions:}
Future research should investigate the specific types of information available in \canvas application READMEs, and whether better descriptions of application functionality as context improve visual bug detection with VLMs.
An additional consideration is how to design new prompting strategies that better target specific \canvas applications.
Also, future research may investigate how \canvas applications vary in visual styles, and how such variances impact the capabilities of VLMs to detect visual bugs.
To address variability in results across \canvas applications, perhaps additional preprocessing of the screenshot or using a VLM fine-tuned for a specific \canvas application would prove effective in improving visual bug detection.

\textbf{VLMs did not always provide consistent visual bug detection results for every repetition of the same experiment.}
While there is inherent randomness in VLM outputs during our experiments, this is a general limitation of leveraging pre-trained deep learning models in software testing.
This is due to the stochastic nature of deep learning models such as VLMs. 
On each run there is no guarantee of which parts of the context the VLM will ``attend'' to, nor what text is generated.

\emph{Future research directions:}
To help address variability in outputs, researchers could seek ways to improve the signal-to-noise (SNR) ratio of context to help ensure VLMs ``attend'' to the (most) relevant parts of the context.
For example, context about how to set up and run the application that may be available in a README could ``distract'' the VLM from the relevant information (how the application is expected to operate) within the README.
Future research should investigate methods to preprocess context and ensure only the most relevant parts are included.
Additionally, future research should investigate better ways to account for randomness in outputs when leveraging VLMs for software testing other than generating multiple outputs.

\begin{tcolorbox}
\textbf{RQ1 implications for developers:}
Applying a prompting strategy like \promptthree when leveraging VLMs can help developers automatically catch visual bugs.
However, developers should evaluate whether leveraging VLMs is useful for their application on a case-by-case basis, as performance varies across applications.
VLMs are best used to target \State visual bugs, and it is reasonable to expect VLMs to be useful for detecting \Rendering and \Layout bugs.
Developers should anticipate that their manual testing processes will still need to target \Appearance bugs more heavily when compared to the other bug types in this study.
Generating multiple text outputs with the VLM per screenshot may slightly increase visual bug detection rates.
\end{tcolorbox}

\subsection*{RQ2: \rqtwo}

\textbf{The bug-free screenshot was an essential piece of context when leveraging VLMs to detect visual bugs.}
The median precision of bug detection leveraging VLMs increased substantially from between 34\% and 50\% without a \bugfree screenshot to 100\% with the inclusion of a \bugfree screenshot as context.
Due to (small) differences in timing between E2E test script execution or randomness inherent to a \canvas application's functionality, screenshots taken between each E2E test script execution can often contain (correct) visual differences for the same test.
As seen in Figure~\ref{fig:intro_sample}, we only require that the (types of) objects expected to be seen on the \canvas bitmap during execution of E2E test scripts are also represented in each respective \bugfree screenshot.
Therefore, it should be possible to avoid the issue of oracle re-verification that is posed by snapshot testing approaches (see Section~\ref{subsec:detectingvisualbugsbackground})  by leveraging VLMs with a prompting strategy like \promptthree.

\emph{Future research directions:}
Future research should investigate methods for automatically creating E2E test scripts for HTML5 \canvas applications.
Once created, E2E test scripts can be repeatedly executed to collect somewhat similar screenshot and compare them across versions of an application, enabling regression testing with VLMs.
However, a difficulty when creating E2E test scripts for the HTML5 \canvas applications was that existing automated testing frameworks such as \textsc{Playwright} do not have extensive support for recording user actions on the HTML5 \canvas.
Therefore, future research could aim to create quick-but-effective approaches for recording user actions on the \canvas for the purposes of saving developers time when they are initially creating E2E test scripts.
Future research could also investigate how to fully automate the creation of E2E test scripts for HTML5 \canvas applications by also leveraging VLMs using an approach similar to those proposed in prior work for non-\canvas web applications~\cite{koh2024visualwebarena, shahbandeh2024naviqate}.

\textbf{Providing the \canvas application README files was helpful for increasing the accuracy of visual bug detection.}
As can be seen in Figure~\ref{fig:resultsA}, prompting strategy \promptone provides higher accuracy than \promptzero.
Additionally, prompting strategy \prompttwo provides higher bug-free accuracy than \promptzero and \promptone but has lower bug-injected accuracy than \promptone.
While the increase in bug-free accuracy with \prompttwo did come at a cost of lower bug-injected accuracy, such a trade-off is necessary to help make the approach useful in practice.
These results suggest that a description of the \canvas application and a description of potential visual bugs should be provided in conjunction with the \bugfree screenshot, rather than supplying the \bugfree screenshot alone.

\emph{Future research directions:}
Future studies should investigate what information in the README files was specifically useful.
While the results of our ablation study indicated that, for example, information about the correct functionality of the application was helpful in increasing recall, future research is required to better understand how individual pieces of context impact the capabilities of VLMs to detect visual bugs.
Additionally, future research should consider how the length of (i.e., number of tokens in) the README file(s) impacts visual bug detection results.
Similarly for the description of visual bugs, studies might consider changing the wording of bug descriptions without changing their overall meaning (e.g., by extending the descriptions with some more examples) to evaluate the impact this has on visual bug detection.

\begin{figure*}[htbp]
\centering

\setlength{\fboxsep}{0pt}%
\begin{subfigure}[t]{0.6\linewidth}
\centering
\fbox{\includegraphics[width=\textwidth]{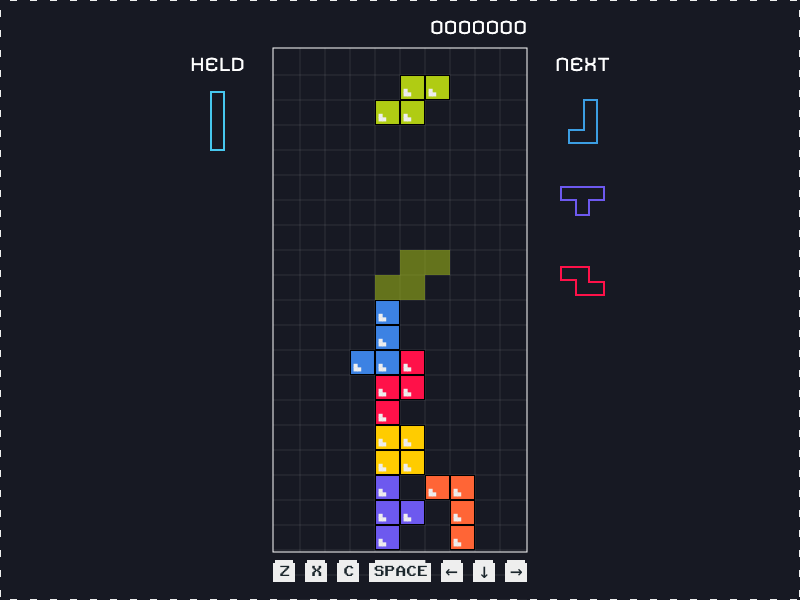}}%
\caption{Screenshot without any visual bugs.}
\label{fig:discussion_a}
\end{subfigure}

\setlength{\fboxsep}{6.25pt}%
\begin{subfigure}[t]{\linewidth}
\fcolorbox{black}{lightgrey}{%
\parbox{0.9\columnwidth}{
{\footnotesize
\textbf{Detected a Visual Bug?} Yes
\\
\textbf{Description of Visual Bug:} \textit{``The green Tetrimino in the center of the playfield appears to be partially transparent or not fully rendered, making it look incomplete. This can disrupt gameplay [...]''}
\\
\textbf{Correct Answer?} No
}}}
\label{fig:discussion_c}
\caption{VLM outputs (\promptzero, \bugfree screenshot).}
\end{subfigure}

\begin{subfigure}[t]{\linewidth}
\fcolorbox{black}{lightgrey}{%
\parbox{0.9\columnwidth}{
{\footnotesize
\textbf{Detected a Visual Bug?} No
\\
\textbf{Description of Visual Bug:} \textit{``''}
\\
\textbf{Correct Answer?} Yes \checkmark
}}}
\caption{VLM outputs (\promptthree, \bugfree screenshot).}
\label{fig:discussion_e}
\end{subfigure}

\setlength{\fboxsep}{0pt}%
\begin{subfigure}[t]{0.6\linewidth}
\centering
\fbox{\includegraphics[width=\textwidth]{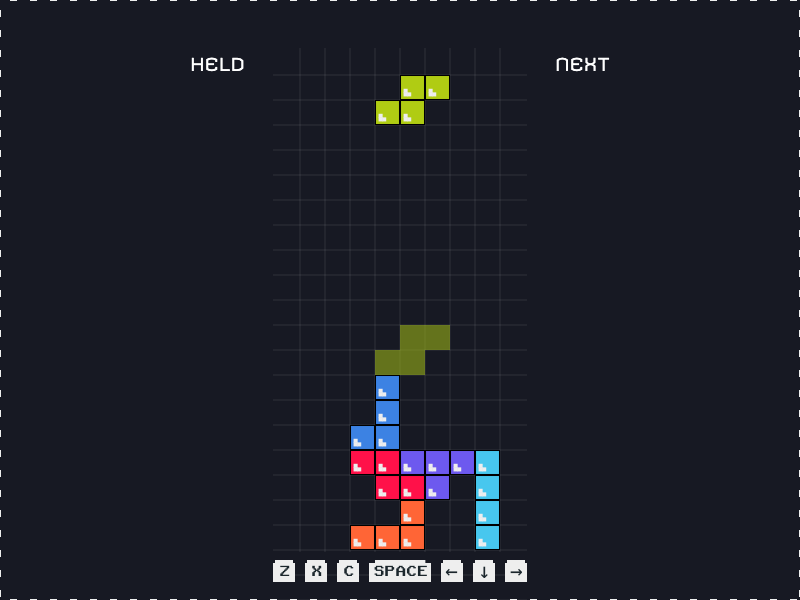}}%
\caption{Screenshot with a \textit{State} bug making the next queue invisible.}
\label{fig:discussion_b}
\end{subfigure}

\setlength{\fboxsep}{6.25pt}%
\begin{subfigure}[t]{\linewidth}
\fcolorbox{black}{lightgrey}{%
\parbox{0.9\columnwidth}{
{\footnotesize
\textbf{Detected a Visual Bug?}  Yes
\\
\textbf{Description of Visual Bug:} \textit{``The green T-piece is floating above the stack and not aligned correctly within the grid, indicating a display or logic issue with the rendering or movement of Tetris pieces.''}
\\
\textbf{Correct Answer?} No
}}}
\caption{VLM outputs (\promptzero, \buginjected screenshot).}
\label{fig:discussion_d}
\end{subfigure}

\begin{subfigure}[t]{\linewidth}
\fcolorbox{black}{lightgrey}{%
\parbox{0.9\columnwidth}{
{\footnotesize
\textbf{Detected a Visual Bug?}  Yes
\\
\textbf{Description of Visual Bug:} \textit{``The ``next'' queue is missing blocks, which might indicate a state bug where upcoming pieces aren't being displayed''}
\\
\textbf{Correct Answer?} Yes \checkmark
}}}
\caption{VLM outputs (\promptthree, \buginjected screenshot).}
\label{fig:discussion_f}
\end{subfigure}

\caption{A bug-injected and a bug-free screenshot from the same \canvas application each paired with the VLM outputs when using prompting strategy \promptzero compared to \promptthree.}
\label{fig:discussion}
\end{figure*}

\textbf{When no additional context was provided, VLMs generated text that addressed bug-free and bug-injected screenshots in the same way.}
We find that the \promptzero prompting strategy is not sufficient for leveraging VLMs as visual bug detectors for HTML5 \canvas applications.
Typically, if no additional context was provided, the VLM produced text that addressed \buginjected and \bugfree screenshots in a very similar way, often focusing on irrelevant aspects of the screenshot or indicating a need for more context.
Even when a visual bug was detected while using the baseline prompting strategy (\promptzero), the VLM output text typically contained an inaccurate description of the bug and often produced a false positive for the \bugfree screenshot from the same application.
Figure~\ref{fig:discussion} shows an example of a \bugfree screenshot and \buginjected screenshot from the same \canvas application, each paired with both the (incorrect) bug detection results when using prompting strategy \promptzero and the (correct) bug detection results when using prompting strategy \promptthree.

\emph{Future research directions:}
Future studies should address whether fine-tuning a VLM with data specific to \canvas applications produces a VLM that does not have the requirement to provide additional information as context when prompting the VLM for visual bug detection.

\textbf{Providing assets did not help for detecting visual bugs when leveraging VLMs.}
Providing samples of image assets did not help when leveraging VLMs for detecting visual bugs in HTML5 \canvas applications.
One potential reason that providing image assets was not helpful in our experiments is that providing the assets may ``confuse'' the VLM about what the correct display of objects looks like, especially as assets are usually transformed (e.g., rasterisation, resizing, tinting) before being displayed on the \canvas.

\emph{Future research directions:}
Future studies should consider how to better leverage the image assets as context.
In our prior work~\cite{macklon2022automatically}, we used the image assets as visual test oracles by applying image transformations determined by information available in the COR.
Therefore, future studies might consider how to perform similar types of transformations to make the image assets more useful as context for the VLM.

\textbf{Using the bug-free screenshot as context did not guarantee 100\% bug-free accuracy.}
The median \bugfree accuracy for prompting strategy \promptthree was 98\%.
Given that experiments evaluating \bugfree accuracy for prompting strategy \promptthree are providing the exact same screenshot as both the \bugfree screenshot for context and as the screenshot for analysis, it is surprising that the median \bugfree accuracy is not instead 100\%.
These results indicate that even when we provided the VLM with the exact same bug-free screenshot, the VLM outputs did not discern that the bug-free screenshot was in fact bug-free 100\% of the time.

\emph{Future research directions:}
While not catastrophic, ideally leveraging VLMs to detect visual bugs would attain 100\% bug-free accuracy with a prompting strategy like \promptthree.
Future studies could consider investigations on what prevents the VLM from generating text that correctly identifies equivalent screenshots, and whether such an issue can be mitigated.

\begin{tcolorbox}
\textbf{RQ2 implications for developers:}
Developers should ensure they have created a README file that helps to explain what the \canvas application is and how it is supposed to function, e.g., what features are part of the application. 
This explanation seems to be particularly helpful for \canvas applications because they are not photorealistic (unlike the video games studied in prior work~\cite{taesiri2024glitchbench}).
Additionally, the requirement for a \bugfree screenshot emphasizes that leveraging VLMs to detect visual bugs has high potential to be used for regression testing.
Screenshots taken from a version of the application known to be \bugfree can be used as the \bugfree screenshot in prompting strategy \promptthree.
\end{tcolorbox}

\section{Threats to Validity} \label{sec:threats}

\subsubsection*{Construct Validity}
We constructed our dataset by injecting visual bugs into 20 FOSS HTML5 \canvas applications collected from GitHub, providing a threat to construct validity related to whether the injected visual bugs were representative of real visual bugs that are detected by human testers in HTML5 \canvas applications.
While the causes of our injected visual bugs may not necessarily be the same as real visual bugs, the visual effects observed are very similar to real visual bugs observed on the \canvas as described in prior work~\cite{macklon2022automatically, macklon2023taxonomy}.
A single author manually verified that the \bugfree screenshots did not contain any visual bugs and that the \buginjected screenshots displayed a visual bug that was straightforward for a human to detect. 
Given that our injected visual bugs are representative of real visual bugs and are straightforward for a human tester to detect, our study mitigates this threat to construct validity.

\subsubsection*{Internal Validity}
To evaluate our experiments, a single author manually checked that the bug descriptions generated by the VLM were correct, providing a threat to internal validity related to coder bias.
In addition to using the available ground truth labels, a single author manually checked each screenshot while manually evaluating the descriptions of the visual bugs that were generated by the VLM, meaning that our results may have been biased when making judgments about the correctness of a description.
We mitigated bias during this process because we already had ground truth labels for the dataset of 100 screenshots that we used in our experiments, and it was simple to determine if the generated description matched the observed bug because the injected visual bugs were straightforward for a human tester to detect.

The methodology we followed to construct a dataset of 100 screenshots contains multiple processes in which a single author manually labelled (or filtered) data.
A threat to validity is coder bias in these manual steps, with one impact being that we may have missed some mature, modern \canvas applications that are built with \textsc{PixiJS}.
We mitigated this threat to validity by considering a wide-variety of information available in each GitHub repository such as its README file while also leveraging tools such as GitHub Search, rather than relying solely on the repository's GitHub description.

\subsubsection*{External Validity}
We set a threshold of 11 GitHub stars to remove toy GitHub repositories from our study~\cite{maj2024fault, munaiah2017curating, kalliamvakou2014promises}.
However, setting such a threshold means that our results may not generalize to the full population of PixiJS dependents~\cite{maj2024fault}, and our study may miss some mature repositories~\cite{munaiah2017curating}.
Future research is required to understand how our results apply to dependents of PixiJS that had less than 11 stars.

The results reported in our study are based on a collection of 100 screenshots taken from a set of 20 FOSS HTML5 \canvas applications that use the \PixiJS framework at major version \textsc{v6} or \textsc{v7} with the WebGL API-based \PixiJS renderer.
Due to the limitations of the dataset used in our experiments, future work is required to better understand how VLMs can be leveraged to detect visual bugs in other applications that do not use \PixiJS \textsc{v6} or \textsc{v7}.

The \buginjected screenshots in our experiments were limited to just a single bug instance per bug type because we used one \buginjected shader program for each bug type across all applications, providing a threat to validity that our results may not extend to other visual bug instances.
Future studies should investigate if our results extend to other specific visual bug instances that are not represented in our dataset.
To support such future research, we release our custom framework (see Section~\ref{sec:introduction}) that can be used to create new \buginjected screenshots by simply creating new \buginjected shaders.

\section{Conclusion} \label{sec:conclusion}
By utilizing a prompting strategy that provides sufficient context about the correct functionality of an HTML5 \canvas application, VLMs can detect visual bugs in HTML5 \canvas applications with relatively high accuracy.
VLMs are most useful for detecting \State bugs, but also can be used to detect a decent proportion of \Rendering and \Layout bugs.
While it is most challenging to detect \Appearance bugs, it is still worthwhile to target all four bug types in our study with a prompting strategy like \promptthree.
The capabilities of VLMs to detect visual bugs can vary widely depending upon the application under test.
Although VLMs are not ready to replace manual testing, they can be useful to reduce the time required during manual testing by catching some of the most obvious visual bugs automatically, enabling manual testers to focus their efforts on detecting a smaller subset of issues in HTML5 \canvas applications.
Overall, leveraging VLMs to detect visual bugs in \canvas applications shows great potential. Future research can leverage our dataset and framework as the first steps towards better ways to leverage VLMs for detecting visual bugs.

\section*{Data Availability Statements}
The datasets generated and analyzed during the current study are available as part of our replication package on Zenodo at the following link:

\url{https://zenodo.org/records/14642612}.

\section*{Compliance With Ethical Standards Statements}
\paragraph*{}
\emph{Conflict of interest:} The authors declare that they have no conflict of interest.
\paragraph*{}
\emph{Funding:} The research reported in this article has been supported by an Alberta Innovates Graduate Student Scholarship (Project ID: 222300640).
\paragraph*{}
\emph{Ethical approval:} N/A.
\paragraph*{}
\emph{Informed consent:} N/A.
\paragraph*{}
\emph{Author contributions:}  Finlay~Macklon was responsible for developing the ideas, collecting open-source \canvas applications, injecting visual bugs into and collecting screenshots from the \canvas applications, designing the prompting strategies used in experiments, conducting experiments, analyzing the results, and writing the manuscript.
Dr.~Cor-Paul~Bezemer was the supervisory author and was involved in concept formation and manuscript composition.

\printbibliography


\newpage

\appendix
\section*{Appendix}\label{appendix}

\subsection*{Prompts used in experiments}

\subsubsection*{\promptzero}

\emph{Message sent to VLM:}
\begin{lstlisting}
Does this screenshot from an HTML5 Canvas application display any visual bugs? If so, please describe the visual bug.
\end{lstlisting}

\subsubsection*{\promptone (also \promptablationA and \promptablationB for ablation study)}

\emph{Message sent to VLM:}
\begin{lstlisting}
Does this screenshot from an HTML5 Canvas application display any visual bugs? If so, please describe the visual bug.

The application has the following README:

{README}
\end{lstlisting}

\subsubsection*{\prompttwo}

\emph{Message sent to VLM:}
\begin{lstlisting}
# Detecting visual bugs in screenshots of HTML5 Canvas applications

## Task description

Please explain if there are any visual bugs in the following screenshot from an application. 
If no visual bug is present, say so. 
If you are unsure, say so.

## Visual bug categories description

The following visual bugs are possible, but may not necessarily be present:

Rendering - Objects appear blurry, distorted, or contain artifacts.
Layout - Objects have incorrect positioning, layering, or size.
State - Objects displayed in the wrong state, e.g. visible vs. invisible.
Appearance - Objects have incorrect aesthetics, e.g. wrong colour.

## Application description

The application has the following README:

{README}
\end{lstlisting}

\subsubsection*{\promptthree}

\emph{Mocked message sent to VLM with bug-free screenshot (response pre-determined):}
\begin{lstlisting}
# Detecting visual bugs in screenshots of HTML5 Canvas applications

## Task description

Please explain if there are any visual bugs in the following screenshot from an application. 
If no visual bug is present, say so. 
If you are unsure, say so.

## Visual bug categories description

The following visual bugs are possible, but may not necessarily be present:

Rendering - Objects appear blurry, distorted, or contain artifacts.
Layout - Objects have incorrect positioning, layering, or size.
State - Objects displayed in the wrong state, e.g. visible vs. invisible.
Appearance - Objects have incorrect aesthetics, e.g. wrong colour.

## Application description

The application has the following README:

{README}
\end{lstlisting}
\emph{Mocked response from VLM (written by human):}
\begin{lstlisting}
This screenshot is free of any visual bugs as defined in the provided set of categories.
\end{lstlisting}
\emph{Message sent to VLM with test screenshot:}
\begin{lstlisting}
Here is another screenshot from the same application. Is there is a visual bug in this screenshot?
\end{lstlisting}

\subsubsection*{\promptfour}

\emph{Mocked message sent to VLM with bug-free screenshot (response pre-determined):}
\begin{lstlisting}
# Detecting visual bugs in screenshots of HTML5 Canvas applications


## Task description

Please explain if there are any visual bugs in the following screenshot from an application. 
If no visual bug is present, say so. 
If you are unsure, say so.


## Visual bug categories description

The following visual bugs are possible, but may not necessarily be present:

Rendering - Objects appear blurry, distorted, or contain artifacts.
Layout - Objects have incorrect positioning, layering, or size.
State - Objects displayed in the wrong state, e.g. visible vs. invisible.
Appearance - Objects have incorrect aesthetics, e.g. wrong colour.


## Application image assets

The Canvas application's image assets have been included in this message as oracles for comparison with the provided screenshot.


## Application description

The application has the following README:

{README}
\end{lstlisting}
\emph{Mocked response from VLM (written by human):}
\begin{lstlisting}
This screenshot is free of any visual bugs as defined in the provided set of categories.
\end{lstlisting}
\emph{Message sent to VLM with test screenshot:}
\begin{lstlisting}
Here is another screenshot from the same application. Is there is a visual bug in this screenshot?
\end{lstlisting}

\subsubsection*{Answer extraction (using structured outputs)}

\emph{Message sent to VLM}
\begin{lstlisting}
The following message describes what was observed in a screenshot from an HTML5 Canvas application.\nPlease fill in the provided JSON schema by determining if the above message indicates whether or not there is a visual bug (`bool_did_detect_visual_bug`), and if so, please provide a summarized description of the detected visual bug (`string_description_of_visual_bug`) based on the following message. If there is no visual bug, please fill the `string_description_of_visual_bug` field with an empty string.
\end{lstlisting}
\emph{JSON schema sent with message to VLM}
\begin{lstlisting}
{
    "type": "json_schema",
    "json_schema": {
        "name": "answer_extraction_response",
        "strict": True,
        "schema": {
            "type": "object",
            "properties": {
                "bool_did_detect_visual_bug": {"type": "boolean"},
                "string_description_of_visual_bug": {"type": "string"}
            },
            "required": ["bool_did_detect_visual_bug", "string_description_of_visual_bug"],
            "additionalProperties": False
        }
    }
}
\end{lstlisting}


\end{document}